\RequirePackage{ifpdf}
\documentclass[12pt,letterpaper]{JHEP3}         % For preprint
\usepackage{epsfig}
\def\be{\begin{eqnarray}}
\def\ee{\end{eqnarray}}
\newcommand{\nn}{\nonumber}
\newcommand\para{\paragraph{}}

\newcommand{\eqn}[1]{(\ref{#1})}

\newcommand\ks[2]{\xi_{(#1)}^{\dot{#2}}}

\def\Dslash{\,\,{\raise.15ex\hbox{/}\mkern-12mu D}}
\def\Dbarslash{\,\,{\raise.15ex\hbox{/}\mkern-12mu {\bar D}}}
\def\delslash{\,\,{\raise.15ex\hbox{/}\mkern-9mu \partial}}
\def\delbarslash{\,\,{\raise.15ex\hbox{/}\mkern-9mu {\bar\partial}}}
\def\pslash{\,\,{\raise.15ex\hbox{/}\mkern-9mu p}}
\def\calDslash{\,\,{\raise.15ex\hbox{/}\mkern-12mu {\cal D}}}

\newcommand{\Tr}{{\rm Tr}}

\newcommand{\reg}{M_{UV}}
\newcommand{\pv}[1]{\left[ #1 \right]_{\rm PV}}

\def\lae{\mathrel{\mathop{\smash{\lower .5 ex \hbox{$\stackrel<\sim$}}}}}
\def\lae{\mathrel{\mathop{\smash{\lower .5 ex \hbox{$\stackrel>\sim$}}}}}

\title{Quantum Dynamics of Supergravity  on ${\bf R}^{3}\times {\bf S}^1$}

\author{David Tong and Carl Turner\\
Department of Applied Mathematics and Theoretical Physics, \\
University of Cambridge, \\
Cambridge, CB3 0WA, UK\\{\tt d.tong, c.p.turner@damtp.cam.ac.uk}}

\abstract{We study the quantum dynamics of ${\cal N}=1$ supergravity in four dimensions with a compact  spatial circle. Supersymmetry ensures that the perturbative contributions to the Casimir energy on the circle cancel. However, instanton  contributions remain. These render  supersymmetric compactification on a circle unstable and the background dynamically decompactifies back to four dimensions. The calculation provides a testing ground for some old ideas in Euclidean quantum gravity.  In particular, we show that gravitational instantons are associated to a new, infra-red scale which is naturally exponentially suppressed relative to the Planck scale and arises from the logarithmic running of the Gauss-Bonnet term. There are also some interesting technical details, including the non-cancellation of  bosonic and fermionic determinants around the background of a self-dual gravitational instanton, despite the existence of supersymmetry. }

\begin{document}
\pagestyle{plain} \setcounter{page}{1}
\newcounter{bean}
\baselineskip16pt \setcounter{section}{0}

\newpage

\section{Introduction and Summary}

The purpose of this paper is to study  four-dimensional  ${\cal N}=1$ supergravity compactified on a spatial circle. We will show that this background is quantum mechanically unstable: the circle dynamically expands and the ground state is Minkowski space with all three spatial dimensions non-compact.

\para
Quantum mechanical instabilities of Kaluza-Klein compactifications have a long history. In the absence of supersymmetry, a Casimir force is generated perturbatively with a competition between bosonic fields, which cause the circle to contract, and fermionic fields which cause the circle to expand \cite{chodos}. More scary instabilities lurk at the non-perturbative level, with  space teetering on the brink of  tunnelling into a bubble of nothing \cite{witten}.

\para
The existence of supersymmetry removes both instabilities described above. But another remains. As we show in some detail, a Casimir force is now generated by gravitational instantons. This results in a  superpotential which schematically takes the form
\be {\cal W} \sim \exp\left(- \frac{\pi R^2}{4G_N} - i\sigma\right)\label{w}\ee
%
%
%\be {\cal W} \sim e^{-2\pi^2 M_{\rm pl}^2 R^2 - i\sigma}\label{w}\ee
%
where $R$ is the radius of the spatial circle and $\sigma$ is dual to the Kaluza-Klein photon, $d\sigma \sim {}^\star F$. The existence of the superpotential \eqn{w} was first proposed in \cite{grimm} on the basis of fermi zero mode counting. It is also closely related to the superpotentials arising from D6-brane instantons wrapping G2-holonomy manifolds described in \cite{hmoore}. Our goal in this paper is to develop the full quantum supergravity computation which results in \eqn{w}. 

\para
 One motivation for performing  the instanton calculation in some detail is that ${\cal N}=1$ supergravity offers a testing ground in which some of the old ideas of Euclidean quantum gravity can be explored, but where many of the accompanying difficulties  do not arise. It thus provides an  opportunity for precision Euclidean quantum gravity. Indeed, as we will see, we will be able to compute the numerical prefactor in \eqn{w}. 
 %Indeed, the Casimir force is non-local from the four-dimensional perspective which ensures that it can be computed  independently of any ultra-violet completion of the theory. 
 In doing  these calculations, we met a number of issues that were (at least to us) surprising and we think worth highlighting. 
 
\subsubsection*{The Scale of Gravitational Instantons}

The natural energy scale associated to any quantum gravity effect is usually thought to  lie far in the ultra-violet, whether Planck scale, string scale or something else. However, in situations where gravitational instantons play a role, this is not the only  scale of importance. The partition function for quantum gravity comes equipped with a hidden infra-red scale, $\Lambda_{\rm grav}$.  This arises through dimensional transmutation from the logarithmic running of the coefficient $\alpha(\mu)$ of the Gauss-Bonnet term,
\be \Lambda_{\rm grav}  = \mu \exp\left(-\frac{\alpha(\mu)}{2\alpha_1}\right)\nn\ee
Here $\alpha_1$ is an appropriate beta-function for the Gauss-Bonnet term.  Of course, gravity is not a renormalisable theory and so, in some sense, includes an infinite number of extra scales associated to the higher-derivative operators. These are all ultra-violet scales, naturally of order of the Planck mass or other UV cut-off. In contrast, the scale  $\Lambda_{\rm grav}$ is distinguished by the fact that, like its Yang-Mills counterpart $\Lambda_{\rm QCD}$,  it can be naturally exponentially suppressed relative to the Planck scale. 

\para
The Gauss-Bonnet term is topological and  the scale $\Lambda_{\rm grav}$ plays no role in perturbative physics around flat space. However, it becomes important when summing over gravitational instantons with non-trivial topology. Moreover, 
in supersymmetric theories, $\Lambda_{\rm grav}$ is naturally complex, with the phase supplied by the gravitational theta angle.  The complexified $\Lambda_{\rm grav}$  lives in a chiral multiplet and, indeed, we  will see that it provides (part of) the pre-factor for the superpotential \eqn{w}. A discussion of this new scale can be found in Section \ref{sumpertsec} and \ref{heatkernelsec}.

\subsubsection*{Summing over Topologies}

One conceptual issue that arises  in this paper is the question of what topologies we should include in the path integral.  We are interested in physics on ${\cal M}\cong{\bf R}^{1,2}\times {\bf S}^1$. In Euclidean space, this manifold has boundary $\partial{\cal M}\cong{\bf S}^2\times {\bf S}^1$. However, the gravitational instantons that we meet have boundaries with different topologies. They are the multi-Taub-NUT spaces, whose boundary is isomorphic to the Lens space $L_k$ in which the ${\bf S}^1$ is non-trivially fibered over the ${\bf S}^2$ with winding $k$. We will argue that we should, nonetheless, include these in the path integral.. The superpotential \eqn{w} arises from the simplest Taub-NUT space in which the ${\bf S}^1$ winds once around ${\bf S}^2$. 
 
 \para
 There are further gravitational instantons whose boundary has the topology of ${\bf S}^1$ fibered over  ${\bf RP}^2\cong {\bf S}^2/{\bf Z}_2$. The Atiyah-Hitchin manifold falls in this class and has the right number of zero modes to contribute to the superpotential.  However, we argue that this class of solutions should be discarded. This discussion can be found in Section \ref{gravinssec}.
 
\subsubsection*{One-Loop Determinants}

The final issue that we wish to highlight is of a more technical nature. In any instanton calculation, one should compute the one-loop determinants around the background of the classical solution. In supersymmetric theories, there is a pairing between the bosonic and fermionic non-zero modes and, correspondingly, a naive expectation that these determinants should cancel. However, for non-compact spaces such as Taub-NUT, the spectrum of operators is continuous and although the range of bosonic and fermionic eigenvalues coincides, their densities need not. We will show that the resulting determinants in Taub-NUT indeed do not cancel but, nonetheless, are computable. They are closely related to the boundary terms that appear in index theory. These determinants are computed in Section \ref{instdetsec}.

\para
The one-loop determinants contribute to the pre-factor of \eqn{w}. Ignoring numerical factors, the superpotential is more precisely given by 
\be {\cal W} \sim \Lambda_{\rm grav}^{41/24}\,R^{-7/24} \exp\left(-\frac{\pi R^2}{4G_N} - i\sigma\right)\nn\ee
The presence of a power of $R$ in the pre-factor appears to be in tension with the expected holomorphy of the superpotential. We will, however, find that there is a one-loop correction to the complex structure relating $R$ and $\sigma$ and that the superpotential above is indeed holomorphic as expected. This discussion can be found in  Section \ref{susysec}.

\subsubsection*{The Plan of the Paper}

We begin in  Section \ref{classec} by describing a few simple classical aspects of  ${\cal N}=1$ supergravity and its Kaluza-Klein compactification to three dimensions.
Section \ref{pertsec} is devoted to perturbative aspects. We start  with a summary of the most important results, including the one-loop divergences that give rise to the new scale $\Lambda_{\rm grav}$, as well as the finite renormalisation of the kinetic terms. The remainder of Section \ref{pertsec} describes these calculations in more detail.  Section \ref{instsec} covers  the instanton computation. We again start with a summary, focussing in particular on the gravitational instantons of interest and a discussion of the kind of asymptotic boundaries that we should admit. The majority of Section \ref{instsec} is concerned with the computation of the one-loop determinants around the background of Taub-NUT.

\para
Readers who would like to skip the gruesome calculational details can get by with reading Section \ref{classec}, Section \ref{sumpertsec} and Section \ref{gravinssec}, before skipping to the punchline at the end.

\section{Classical Aspects}\label{classec}

We work with ${\cal N}=1$ supergravity in $d=3+1$ dimensions. Throughout the paper, we focus on the minimal theory containing only a graviton and gravitino. The  bulk four-dimensional action is given by
\be S = \frac{M_{\rm pl}^2}{2}\int d^4x\,\sqrt{-g}\left({\cal R}_{(4)} + \bar{\psi}_\mu\gamma^{\mu\nu\rho}{\cal D}_\nu \psi_\rho\right)\label{readyset}\ee
We use the notation of the (reduced) Planck mass $M_{\rm pl}^2 = 1/8\pi G_N$ instead of the Newton constant $G_N$. 
Here ${\cal R}_{(4)}$ is the 4d Ricci scalar, with the subscript  to distinguish it from its 3d counterpart that we will introduce shortly. There is also the standard Gibbons-Hawking boundary term which we have not written explicitly.

\para
The action is to be thought of as a functional of the Majorana gravitino $\psi_\mu$ and the vierbein $e^a_\mu$ where $\mu,\nu=0,1,2,3$ are spacetime indices and $a,b=0,1,2,3$ are tangent space indices. Here we follow the standard notation of suppressing the spinor indices on the gravitino, whose covariant derivative is given by
\be {\cal D}_\nu \psi_\rho = \partial_\nu\psi_\rho + \frac{1}{4}\hat{\omega}_{ab\nu}\gamma^{ab}\psi_\rho\nn\ee
In this formalism, the spin connection $\hat{\omega}_{ab\mu}$ that appears in the covariant derivative differs from the purely geometric spin connection by the addition of a gravitino torsion term: $\hat{\omega}_{ab\mu} = \omega_{ab\mu}(e) + H_{ab\mu}$ with
\be H_{ab\mu} = -\frac{1}{4} e^\nu_a e^\rho_b\left( \bar{\psi}_\mu \gamma_\rho \psi_\nu - \bar{\psi}_\nu \gamma_\mu \psi_\rho - \bar{\psi}_\rho \gamma_\nu \psi_\mu\right)\nn\ee
The action is, of course, invariant under diffeomorphisms and local supersymmetry transformations. The latter act as $\delta e_\mu^a = \frac{1}{2}\bar{\epsilon}\gamma^a\psi_\mu$ and $\delta\psi_\mu= {\cal D}_\mu\epsilon$. 

\para
The classical theory  also enjoys a $U(1)_R$ symmetry which acts by axial rotations on $\psi$. As we describe in more detail in Sections \ref{pertsec} and \ref{instsec}, this $U(1)_R$ symmetry is  anomalous in the quantum theory. (Although, as we will see, it mixes with a $U(1)_J$ bosonic symmetry that will be described shortly and a combination of the two survives.) 

\subsection{Reduction on a Circle}

Our interest in this paper is in the dynamics of ${\cal N}=1$ supergravity when compactified on a manifold ${\cal M}\cong {\bf R}^{1,2}\times {\bf S}^1$. We denote the physical radius of the circle as $R$. We choose the spin structure such that the fermions are periodic around the compact direction and supersymmetry is  preserved.

\para
At distances larger than the compactification scale $R$, the dynamics is effectively three dimensional. The metric degrees of freedom are parameterised by the familiar Kaluza-Klein ansatz,
\be ds_{(4)}^2 = \frac{L^2}{R^2}ds_{(3)}^2 + \frac{R^2}{L^2}\left(dz + A_i dx^i\right)^2\label{kkansatz}\ee
where $z\in [0,2\pi L)$ is the periodic coordinate. Here $R$, $A_i$ and the 3d metric $g^{(3)}_{ij}$ are dynamical degrees of freedom, while $L$ is a fixed, fiducial scale. It is natural to pick coordinates such that $R(x)\rightarrow L$ asymptotically and we will eventually do so but, for now, we leave $L$ arbitrary.

\para
Evaluated on this background, the Einstein-Hilbert action becomes
\be S_{\rm eff} &=& \frac{M_{\rm pl}^2}{2}\int d^4x\,\sqrt{-g}\,{\cal R}_{(4)} \nn\\ &=& \frac{ M_3}{2}\int d^3x \,\sqrt{-g_{(3)}}\left[{\cal R}_{(3)} -2\left(\frac{\partial R}{R}\right)^2 -\frac{1}{4}\frac{R^4}{L^4}F_{ij}F^{ij}\right]
\nn\ee
with $M_3 = 2\pi L M_{\rm pl}^2$ the 3d Planck scale and  $F_{ij} = \partial_iA_j-\partial_j A_i$ the graviphoton field strength.

\para
In three dimensions, it is often useful to dualise the gauge field in favour of a periodic scalar $\sigma$. This is particularly true if we are interested in instanton physics \cite{polyakov}.  The dual photon can be viewed as  Lagrange multiplier which imposes the Bianchi identity,
\be {\cal L}_{\rm \sigma} = \frac{\sigma}{4\pi L} \epsilon^{ijk} D_{i}F_{jk}\label{sigmacoupling}\ee
With the  magnetic charge  quantised in integral units,  $\sigma$ has periodicity $2\pi$. Integrating out the field strength, we can write the low-energy effective action in dual form,
\be S_{\rm eff} = \int d^3x \,\sqrt{-g_{(3)}}\,\left[\frac{M_3}{2}\,{\cal R}_{(3)} - M_3\left(\frac{\partial R}{R}\right)^2 - \frac{1}{M_3} \frac{L^2}{R^4}\left(\frac{\partial \sigma}{2\pi}\right)^2
\right]
\label{clasact}\ee
This action enjoys a new $U(1)_J$ symmetry which acts by shifting the dual photon: $\sigma\rightarrow \sigma +c$. All other fields are left invariant under this symmetry. The symmetry is preserved in perturbation theory but, as we will see in Section \ref{instsec}, is broken by instanton effects.

\para
Our goal in this paper is to determine the quantum corrections to the effective action \eqn{clasact}. We describe perturbative corrections in Section \ref{pertsec} and instanton corrections in Section \ref{instsec}.

\subsubsection*{Fermions}

This bosonic effective action has a fermionic counterpart which is dictated by supersymmetry. Let us work for now with a Majorana basis of 4d gamma matrices,
\be \gamma^i = \left(\begin{array}{cc} 0 & \gamma_{3d}^i \\ \  \gamma_{3d}^i & \ 0\end{array}\right)\ \ \ i=0,1,2\ \ \ ,\ \ \ \gamma^z = \left(\begin{array}{cr} 1 & \ 0 \\ 0 & \ -1\end{array}\right) \label{majgamma}\ee
with $\gamma_{3d}^i=(i\sigma^2,\sigma^3,\sigma^1)$. 
Upon dimensional reduction, the 4d Majorana gravitino $\psi_\mu$ decomposes into a 3d spin-3/2 Dirac fermion $\lambda_i$ and 3d spin-1/2 Dirac fermion $\chi$.  To perform this reduction, it's simplest to work with the frame index, so that $\psi_a = e_a^\mu \psi_\mu$. Further, to make life easy for ourselves, we restrict to the flat background ${\bf R}^{1,2}\times {\bf S}^1$ with metric \eqn{kkansatz} and make the spinor ansatz, 
\be \psi_i = \left(\begin{array}{c}{\rm Re}\lambda_i +(\gamma_{3d})_i {\rm Im}\chi \\ {\rm Im}\lambda_i + (\gamma_{3d})_i {\rm Re}\chi\end{array}\right) \ \ \ {\rm and}\ \ \ \psi_z = \left(\begin{array}{c} {\rm Re}\chi\\{\rm Im}\chi\end{array}\right)\label{fermidimred}\ee
The gravitino kinetic term in \eqn{readyset} then becomes,
\be S_{\rm fermions} &=&\int d^4x\,\sqrt{-g}\ \frac{M_{\rm pl}^2}{2}\,\bar{\psi}_\mu\gamma^{\mu\nu\rho}{\cal \partial}_\nu \psi_\rho \nn\\ &=& \int d^3x\,\sqrt{-g_{(3)}}\ \frac{M_3L}{R}\left(\frac{1}{2}\bar{\lambda}_i\epsilon^{ijk}\partial_i\lambda_k -\bar{\chi}\!\delslash\chi\right)\label{chikin}\ee
After dividing out by local supersymmetry transformations, the spin-3/2 fermion $\lambda_i$ carries no propagating degrees of freedom. (This is the supersymmetric analog of the statement that the 3d metric carries no propagating degrees of freedom.) In contrast, the spin-1/2 fermion $\chi$ carries two propagating degrees of freedom; these are the supersymmetric partners of $R$ and $\sigma$. We will postpone a  more detailed discussion of how supersymmetry relates $R$, $\sigma$ and $\chi$ to Sections \ref{susysec} and \ref{finalsec}.

\subsection{Topological Terms}

In addition to the Einstein-Hilbert action, there are two topological terms that will play a  role in our story. Both are higher derivative terms, with dimensionless coefficients. They are the Gauss-Bonnet term
\be 
S_\alpha = \frac{\alpha}{32\pi^2} \int d^4x\sqrt{g}  \ {}^\star{\cal R}^\star_{\mu\nu\rho\sigma}\,{\cal R}^{\mu\nu\rho\sigma}
\label{gb}\ee
which integrates to the Euler characteristic of the manifold, and the Pontryagin class,
\be
S_\theta = \frac{\theta}{16\pi^2}\int d^4x\sqrt{-g}  \  {}^\star{\cal R}_{\mu\nu\rho\sigma}{\cal R}^{\mu\nu\rho\sigma}\label{pont}\ee
If we care only about perturbative physics on ${\bf R}^3\times {\bf S}^1$, then we can neglect these terms. However, when we start to sum over manifolds of different topology, they become important. 

\para
Usually, when working with an effective field theory, we keep all relevant and marginal terms in the action, neglecting only the irrelevant operators on the grounds that they are suppressed by some high mass scale. In the present case, there are  two further four-derivative terms which come with dimensionless coefficients:  ${\cal R}^2$ and ${\cal R}_{\mu\nu}{\cal R}^{\mu\nu}$. However, both can be absorbed into the Einstein-Hilbert term through a redefinition of the metric \cite{tv}. For this reason, we need only consider $S_\alpha$ and $S_\theta$ above.

\para
In supergravity, the Gauss-Bonnet \eqn{gb}  and Pontryagin \eqn{pont} terms can be written as an F-term  \cite{paul,ferrara} (using the so-called ``chiral projection operator"). This, in turn, means that the two coupling constants $\alpha$ and $\theta$ combine into the complex coupling
\be \tau_{\rm grav} = \alpha + 2i\theta\label{taugrav}\ee
which naturally lives in a chiral multiplet. We will see later that $\tau_{\rm grav}$ appears in the instanton generated superpotential.

\section{Perturbative Aspects}\label{pertsec}

In this section we describe the results of quantum fluctuations of the graviton and gravitino around the background ${\bf R}^{1,2}\times {\bf S}^1$. There are two kinds of effects: those from divergences that arise already in four dimensions; and finite corrections to the low-energy effective action which are suppressed by the dimensionless combination $1/M_{\rm pl}
^2R^2$. 

\subsection{Summary}\label{sumpertsec}

We open this section by summarising the main results. The remainder of the section contains details of the computations.

\subsubsection*{Finite Corrections}

Finite corrections to the effective action occur when the theory is compactified on  ${\bf R}^{1,2}\times {\bf S}^1$ and arise due to loops wrapping the spatial circle.  The results depend on $R$, the radius of the circle and so are non-local from the four-dimensional perspective.  For this reason, they are not sensitive to the ultra-violet details of the theory and can therefore be reliably calculated.
  
\para
These finite corrections were first computed in the Kaluza-Klein context in \cite{chodos}, where they manifested themselves as a Casimir force, causing the Kaluza-Klein circle to either shrink or expand. (The analogous calculation was performed earlier in the thermal context \cite{gmy}.) The effective 3d potential is given by\footnote{The standard Casimir potential in four dimensions scales as $1/R^3$. The $1/R^6$ scaling seen here arises after a Weyl transformation to the 3d Einstein frame.}
\be V_{\rm eff} = -\frac{N_B-N_F}{720\pi}\frac{L^3}{R^6}\label{pertcasimir}\ee
 Here $N_B$ is the number of massless bosonic degrees of freedom; these make the Kaluza-Klein circle contract.  $N_F$ the number of massless fermionic degrees of freedom; these make the circle expand. Of course, in supersymmetric theories $N_B=N_F$ and Kaluza-Klein compactifications are perturbatively stable.  The presence of fermions with periodic boundary conditions means that the bubble-of-nothing instability is  absent in this theory \cite{witten}, but other gravitational instantons, discussed in Section \ref{instsec}, will contribute. 
  
\para
Although the perturbative potential vanishes,  there are still finite one-loop effects of interest. These renormalise the kinetic terms in the effective action \eqn{clasact}. Much of this section is devoted to computing these effects; we will  show that the  low-energy effective action  becomes,
\be {\cal L}_{\rm eff}  &=& \frac{1}{2}\left(M_3  + \frac{5}{16\pi}\frac{L}{R^2}\right)\,{\cal R}_{(3)} - \left(M_3 - \frac{1}{6\pi}\frac{L}{R^2}\right)\left(\frac{\partial R}{R}\right)^2 \nn\\ &&\ \ \ \ \ \ \ \ \ \ \ \ \ \ \ \ \ \ \ \ \ \ \ \ \  \ \ \ \ \ \ \ \ \ \ - \left(M_3 + \frac{11}{24\pi}\frac{L}{R^2}\right)^{-1} \frac{L^2}{R^4}\left(\frac{\partial \sigma}{2\pi}\right)^2
\ \ \ \ \label{loopact}\ee
This is the one-loop effective action. We certainly expect that there will be further corrections, both from higher-loops and from non-perturbative effects. Nonetheless, this will suffice for our purposes. The most important fact that we will need is the observation that the renormalisation of the $R$ and $\sigma$ kinetic terms come with different coefficients. This will prove important later when we reconcile this with supersymmetry: it results in a one-loop shift in the complex structure and $R$ and $\sigma$ sit together in a chiral multiplet with lowest component
\be {\cal S} = 2\pi^2 M_{\rm pl}^2 R^2 + \frac{7}{48}\log(M^2_{\rm pl}R^2) +i\sigma\nn\ee
The log term above arises from the one-loop correction.  This will be described in Section \ref{susysec}.

\subsubsection*{Anomalies and One-Loop Divergences}

It is well known that the S-matrix is  one-loop finite in pure Einstein gravity \cite{tv} and two-loop finite in pure ${\cal N}=1$ supergravity \cite{petervan,kay}. Nonetheless, these theories do suffer from divergences at one-loop which, while not appearing in the S-matrix, can affect the physics. As we review, these divergences are related to anomalies.

\para
For our purposes, the most important one-loop divergence is associated to the Gauss-Bonnet term \eqn{gb}.  
This, of course, is a total derivative in four-dimensions but will be important when we come to discuss gravitational instanton physics.  The coefficient $\alpha$ is dimensionless and runs logarithmically at one-loop \cite{tv}
\be\alpha(\mu) = \alpha_0 - \alpha_1 \log\left(\frac{\reg^2}{\mu^2}\right)\label{running}\ee
where $\alpha_0$ is the coupling at the UV cut-off which we denote as $\reg$. In general, for a  theory with $N_s$ free massless spin-$s$ fields, the beta-function is given by \cite{perry,cduff,yoneya}
\be \alpha_1 = \frac{1}{48\cdot 15}\left(848 N_2 - 233 N_{3/2} - 52 N_1 + 7 N_{1/2} + 4N_0\right)\nn\ee
The computation leading to this result is closely related to the trace anomaly for massless fields in fixed, curved spacetime. Indeed, for spins  $s\leq 1$, the coefficients above are the same as $c-a$ of the trace anomaly. The running coupling $\alpha(\mu)$ results in an RG-invariant scale,  
\be \Lambda_{\rm grav}  = \mu \exp\left(-\frac{\alpha(\mu)}{2\alpha_1}\right)\label{lambda}\ee
 For the pure supergravity theory that is our focus in this paper we have $N_0=N_{1/2}=N_1=0$ while $N_{3/2}=N_2=1$ which gives $\alpha_1 = 41/48$. 
 
 \para

In the original discussions of Euclidean quantum gravity, the suggestion seems to have been that $\Lambda_{\rm grav}$ (or sometimes $\mu$) should be identified with the Planck scale. (See, for example, \cite{hawkingbook}.) In contrast, here we view $\Lambda_{\rm grav}$ as a new scale which emerges in quantum gravity through dimensional transmutation; it dictates the length at which topological fluctuations are unsuppressed by the Gauss-Bonnet term.  Like its counterpart $\Lambda_{\rm QCD}$ in Yang-Mills theory, $\Lambda_{\rm grav}$  can naturally be exponentially smaller than the  Planck scale. As we will see shortly, like its Yang-Mills counterpart, it provides the scale at which instanton effects become important.

\para
In the previous section, we saw that $\alpha$ sits in a background chiral multiplet with the gravitational theta-term $\theta$. These combine into the complex coupling $\tau_{\rm grav} = \alpha + 2i\theta$. This means that the scale $\Lambda_{\rm grav} =  \mu e^{-\tau/2\alpha_1}$ is also naturally complex in supergravity and sits in a chiral multiplet.

\para
There is one further one-loop divergence that will play a role in our story. This is responsible for  the axial anomaly for the  $U(1)_R$ symmetry with current  $J^\mu_5 = i \bar{\psi}_\nu\gamma^{\nu\mu\rho}\gamma^5\psi_\rho$.  In general, the anomaly is given by \cite{salam,eguchi,cduff}
\be
\nabla_\mu J^\mu_5 = \frac{1}{24\cdot{16\pi^2}}\left(21 N_{3/2} - N_{1/2}\right)\,{}^\star{\cal R}_{\mu\nu\rho\sigma}\,{\cal R}^{\mu\nu\rho\sigma}\label{anomaly}\ee
For us, $N_{1/2} = 0$ and $N_{3/2}=1$. As usual, the anomaly can be compensated by shifts on the gravitational theta angle which means that we should view $\Lambda_{\rm grav}$ as carrying $U(1)_R$ charge.

%\para
%In supergravity, the Gauss-Bonnet \eqn{gb}  and Pontryagin \eqn{pont} terms sit together as the lowest components of a chiral multiplet \cite{paul,ferrara}. This, in turn, means that the two coupling constants $\alpha$ and $\theta$ combine into the complex coupling
%
%\be \tau_{\rm grav} = \alpha + \frac{i}{2}\theta\label{taugrav}\ee
%
%Both $\tau$ and the complexified strong coupling scale $\Lambda_{\rm grav} = \mu e^{-\tau/2\alpha_1}$ then naturally live in chiral multiplets. In particular, we will see later that $\Lambda$ appears in the instanton-generated superpotential. 

\subsection{One-Loop Determinants}\label{detsec}

In this section, we present the determinants arising from one-loop fluctuations of the graviton, the gravitino and their ghosts. This material is standard fare but, since we will need this for a number of subsequent calculations, we take the time to describe it in some detail.

\subsubsection*{The Graviton and its Ghost}

Throughout this paper, we use the background field method. We work in Euclidean space and write the metric as background $g_{\mu\nu}$, which is taken to obey the Einstein equations,  and fluctuation $h_{\mu\nu}$,
\be g_{\mu \nu }\rightarrow {g}_{\mu\nu} + h_{\mu\nu}\nn\ee
From now on, all covariant derivatives and curvatures are to be thought of with respect to the background. It's useful to further decompose the fluctuations into the trace $h=g^{\mu\nu}h_{\mu\nu}$ and traceless parts, $\bar{h}_{\mu\nu} = h_{\mu\nu} - \frac{1}{4}g_{\mu\nu}h$.

\para
We expand the Einstein-Hilbert action to quadratic order in $h_{\mu\nu}$ following, for
example, \cite{ghp}. The residual gauge freedom $h_{\mu\nu} \rightarrow h_{\mu\nu} +
\nabla_\mu\xi_\nu + \nabla_\nu\xi_\mu$ is fixed by imposing the condition
\be  \nabla^\mu\left(h_{\mu\nu} - \frac{1}{2}g_{\mu\nu} h\right)=0\nn\ee
 The resulting Fadeev-Popov determinants are exponentiated in the usual fashion through
the introduction of ghosts which, in this context, are anti-commuting complex vectors.

\para
The Einstein-Hilbert action is, famously, unbounded below. In the present context, this shows up in
the negative-definite  operator $\nabla^2$ for the trace fluctuations $h$. We follow the prescription of \cite{ghp} and rotate the contour to integrate
over imaginary conformal factors so that we work with the positive definite operator
\be
\Delta_0 =  -\nabla^2\label{delzero}\ee
 For the ghosts and traceless fluctuations, no such rotation is necessary. The operators for these other fields are most conveniently written using tangent space indices. This means, for example, that we write the metric fluctuation as  $h_{ab} = e_{a}^\mu e_b^\nu h_{\mu\nu}=
e^\mu_{(a}\,\delta e^{\,}_{b)\mu}$. (Note that the asymmetric components of $e_\mu^a$ are non-propagating.) The fluctuation operator for the symmetric, traceless spin-2 field 
 $\bar{h}_{ab}$ and is given by
  \be (\Delta_{2})_{ab;cd} = -\frac{1}{4}\eta_{ac}\eta_{bd}\nabla^2 +
\left(\frac{1}{4}\eta_{ac}\eta_{bd}{\cal R} - \frac{1}{2}\eta_{ac}{\cal R}_{bd} -
\frac{1}{2}{\cal R}_{acbd}\right)\label{deltwo}\ee
 Meanwhile the fluctuation operator for the spin-1 ghosts takes the form,
 \be ({\Delta}_{1})_{a;b} = -\eta_{ab} \nabla^2 - {\cal R}_{ab}\label{delone}\ee
 Note that in each of \eqn{delzero}, \eqn{deltwo} and \eqn{delone}, the subscript on $\Delta_s$ labels the spin of the field and therefore determines the appropriate Laplacian $\nabla^2$.
 Integrating out the graviton and its ghost at one-loop then results in the determinant factor,
 \be \Gamma_B= \frac{\det\Delta_{1}}{\det^{1/2}\Delta_{2}\,\det^{1/2}\Delta_0}\label{gravitondet}\ee

\subsubsection*{The Gravitino and its Ghost}

The quantisation of the spin-3/2 fermion was described in \cite{nielsen,kallosh}. (See also \cite{perry}.) We again need to fix the redundancy of local supersymmetry transformations. The standard choice is $\gamma^\mu \psi_\mu=0$. After gauge fixing, the kinetic term for the gravitino reads
\be {\cal L}_{\rm gravitino} = \frac{i}{2}\bar{\psi}_a\,\left(\gamma^b\Dslash\gamma^a\right)\psi_b\nn\ee
In computing the one-loop determinants, it's somewhat simpler to work with the squares of Dirac operators. For the spin-3/2 gravitino, this is given by
\be \left(\Delta_{3/2}\right)_{a;b} = \left(\gamma^c\Dslash\gamma_a\right)\left(\gamma_b\Dslash\gamma_c\right) = -\eta_{ab}\nabla^2 - \frac{1}{2}{\cal R}_{cdab}\gamma^{[c}\gamma^{d]} +{\cal R}_{ab}\label{delthree}\ee
where we have left the Dirac spinor indices implicit in this expression.
Meanwhile, the gravitino is accompanied by three commuting, spin-1/2 Majorana ghosts. These come with the simple Dirac operator $i\Dslash$ which, after squaring, becomes
\be \Delta_{\rm 1/2} = (i\Dslash)^2 = -\nabla^2 + \frac{1}{4}{\cal R}\label{delhalf}\ee
Integrating out the gravitino and its ghosts then give rise to the one-loop determinants
\be \Gamma_{F}= \frac{{\rm det}^{1/4}\Delta_{3/2}}{{\rm det}^{3/4}\Delta_{1/2}}\label{gravitinodet}\ee

\subsubsection*{The One-Loop Effective Action}

Each of the one-loop fluctuation operators introduced above takes the form,
\be \Delta_s = -\nabla^2 - E_s\nn\ee
where, for each spin $s=0,\frac{1}{2},1,\frac{3}{2},2$,  the operator includes a spin-dependent term $E_s$, linear in the curvature ${\cal R}_{abcd}$ and is given, respectively, in \eqn{delzero}, \eqn{delhalf}, \eqn{delone}, \eqn{delthree} and \eqn{deltwo}. Of course, the Laplacian $\nabla^2$ also hides a spin structure since acting on the spin $s$ field,
\be \nabla_\mu = \partial_\mu + \frac{1}{2}\omega_{ab\mu}t^{ab}_{(s)}\nn\ee
where $t^{ab}_{(s)}$ are the spin-$s$ Lorentz generators (or, in Euclidean space, rotation generators). 

\para
The one-loop determinants from gravitons \eqn{gravitondet} and gravitinos \eqn{gravitinodet} can be exponentiated to give the one-loop contribution to the effective action. This can be written as
\be S_{\rm 1-loop} = -\sum_{s=0}^2 \zeta_s\,\log\det\Delta_s\label{ratio}\ee
where the coefficients $\zeta_s$ are the exponents of the various operators, given by
\be \zeta_s = \left(-\frac{1}{2},-\frac{3}{4},+1,+\frac{1}{4},-\frac{1}{2}\right)\ \ \ \ \ \ \ \ s=0,\,\frac{1}{2},\,1,\,\frac{3}{2},\,2\nn\ee
The number of off-shell degrees of freedom of a spin-$s$ field are $d_s = (1,4,4,16,9)$. (Recall that the spin-2 operator acts on  the traceless part of symmetric tensors which is why $d_2=9$.)   Note that $\vec{\zeta}\cdot\vec{d}=0$. This, of course, is the manifestation of supersymmetry in the guise of an equal number of bosonic and fermionic off-shell degrees of freedom.

\para
In the rest of this section, we will compute various terms in the expansion of \eqn{ratio}. We will also return to compute the ratio of determinants $\Gamma_B\Gamma_F$  in Section \ref{instdetsec} in a self-dual background where, as we show, considerable simplifications occur. 

%Putting together the graviton and gravitinos, the one-loop determinants faround a general background are given by the ratio,
%
%\be \Gamma=  \frac{\det\Delta_{\rm ghost}\,{\rm det}^{1/4}\Delta_{\rm gravitino}}{\det^{1/2}\Delta_{\rm
%traceless}\,\det^{1/2}\Delta_0\,{\rm det}^{3/4}\Delta_{\rm spinor}}\label{ratio}\ee
%

\subsection{Two-Derivative Effective Action}\label{gradsec}

We first compute the finite corrections to the low-energy effective that we previewed in \eqn{loopact}. Specifically, we compute the one-loop effective action \eqn{ratio} in a gradient expansion around the flat background ${\bf R}^3\times {\bf S}^1$,  keeping only terms with two derivatives or fewer. As we will see, supersymmetry means that many of the contributions vanish.

\para
We take the flat metric to be given by \eqn{kkansatz} with $A_i=0$ and $R$ constant. We denote this metric as $\hat{g}_{\mu\nu}$ and the associated Laplacian as $\hat{\nabla}^2$. Each of the terms in the low-energy effective action can then be expanded as
\be \log \det \Delta_s 
%&=& \Tr\log [\nabla^2 + (\Delta_s - \hat{\nabla}^2)] \nn\\ 
&=& \Tr\log[-\hat{\nabla}^2] + \Tr\log[1 - \hat{\nabla}^{-2}(\Delta_s+\hat{\nabla}^2)] \nn\\ &\approx& \Tr\log[-\hat{\nabla}^2] + \Tr\,(-\hat{\nabla}^{-2})(\Delta_s+\hat{\nabla}^2)\label{delexpand}\\ &&\ \ \ \ \ \ \ \ \  \ \ \ \ \ \ \ \  -\frac{1}{2}\Tr\,(-\hat{\nabla}^{-2})(\Delta_s+\hat{\nabla}^2)(-\hat{\nabla}^{-2})(\Delta_s+\hat{\nabla}^2) +\ldots\nn\ee
where higher order terms do not contribute to the two-derivative effective action.
The leading term above, involving only $\hat{\nabla}^2$, provides the perturbative contribution to the Casimir energy advertised previously in \eqn{pertcasimir}. For us, supersymmetry ensures  it vanishes after summing over all spins, due to the relation $\vec{\zeta}\cdot \vec{d}=0$.

\para
Subsequent terms in the expansion also enjoy cancellations. To see this, let's start with the second term, $\Tr\,(-\hat{\nabla}^{-2})(\Delta_s+\hat{\nabla}^2)$. Expanding the Laplacian, the general fluctuation operator can be written as
\be \Delta_s  = -g^{\mu\nu}\partial_\mu\partial_\nu - \frac{1}{2}g^{\mu\nu}\{\partial_\mu,\omega_{ab\nu}t^{ab}_{(s)}\} - \frac{1}{4}\omega_{ab\mu}\omega_{cd}^{\ \ \mu} t^{ab}_{(s)}t^{cd}_{(s)} + g^{\mu\rho}{\Gamma}^\nu_{\mu\rho}\nabla_\nu - E_s\ \ \ \ \ \ \label{deltas}\ee
The sum over different spins $s=0,\ldots,2$ will mean that any term which doesn't have an explicit spin dependence will vanish. That immediately kills the $\partial^2$ term and the term with the Christoffel symbol. The term linear in $t^{ab}_{(s)}$ vanishes as soon as the trace over spin indices is taken. We're left with
\be \sum_{s=0}^2\zeta_s\, \Tr\,(-\hat{\nabla}^{-2})(\Delta_s+\hat{\nabla}^2) = \sum_{s=0}^2\zeta_s\,\Tr\,(-\hat{\nabla}^{-2})\left[ - \frac{1}{4}\omega_{ab\mu}\omega_{ab}^{\ \ \mu} t^{ab}_{(s)}t^{cd}_{(s)} - E_s\right] \nn\ee
Here the trace $\Tr$ should be taken over both spin and  momentum quantum numbers. We deal with the spin trace first. We have
\be {\rm tr}_{\rm spin} [t^{ab}_{(s)}t^{cd}_{(s)}] = a_s(-\delta^{ac}\delta^{bd} + \delta^{bc}\delta^{ad})\label{ttstar}\ee
where the coefficients $a_s$ are related to the Casimirs of the representation of the Lorentz group\footnote{The irreducible representation $(j_1,j_2)$ has dimension $d=(2j_1+1)(2j_2+1)$ and the appropriate group theory gives  $a=d/3[(j_1(j_1+1)+j_2(j_2+1)]$. (See, for example, \cite{cduff}.)} and are given by
\be a_s = (0,1,2,12,12)\nn\ee
Meanwhile, the trace over spin indices of $E_s$ is proportional to the Ricci scalar of the background:
\be
{\rm tr}_{\rm spin}\,E_s = -b_s{\cal R}\ \ \ \ \ {\rm with}\ \ \ \ \ b_s = (0,1,-1,4,6) \label{ebs}\nn\ee
This allows us to express the contribution to the one-loop effective action in terms of traces over momentum states only. 
\be  \sum_{s=0}^2\zeta_s\, \Tr\,(-\hat{\nabla}^{-2})(\Delta_s+\hat{\nabla}^2) = \frac{1}{2}(\vec{a}\cdot\vec{\zeta})\, \Tr\,[-\hat{\nabla}^{-2}\omega_{abc}\omega^{abc}] + (\vec{b}\cdot\vec{\zeta})\,\Tr\,[-\hat{\nabla}^{-2}{\cal R}]\ \ \ \  \ \ \label{xtwo}\ee
We'll perform these momentum integrals shortly. But, first, we also need to include the contributions from the third term in \eqn{delexpand}. 
\be X_3=-\frac{1}{2}\sum_{s=0}^2 \,\zeta_s\,\Tr\,(-\hat{\nabla}^{-2})(\Delta_s+\hat{\nabla}^2)(-\hat{\nabla}^{-2})(\Delta_s+\hat{\nabla}^2)\nn\ee
Once again, any term   linear in $t^{ab}_{(s)}$ upon  taking the trace over spin indices,  while any term without a spin structure vanishes after summing over different spins due to supersymmetry. After the dust settles, we find that just two terms are relevant,
%%%%%
%%%%%
%%%%% NOTE: There are two extra terms. The $\omega^2 t^2 E$ term is higher derivative since $\omega$ is one derivative and $E$ is the Riemann tensor which is two derivatives. 
%%%%% More subtle is the $\partial^2 \omega^2 t^2$ term. The $\partial^2$ really comes with a $\Delta g^{\mu\nu}\partial_\mu\partial_\nu$, so this is third order in the fields.
%%%%% Probably this term is needed to change the $(\partial R)^2/R_\infty^2$ that we actually get into $(\partial R)^2/R^2$ which we need. Either way, they can't change
%%%%% the leading order terms that we want. 
%
\be X_3 &=& - \sum_{s=0}^2{\rm Tr} \left[ \frac{1}{2}(-\hat{\nabla}^{-2}) t^{ab}_{(s)}  \omega_{ab}\,^\mu \partial_\mu (-\hat{\nabla}^{-2})  t^{cd}_{(s)} \omega_{cd}\,^\nu \partial_\nu 
+(-\hat{\nabla}^{-2})E_s (-\hat{\nabla}^{-2})\left(g-\hat{g}\right)^{\mu\nu} \partial_\mu \partial_\nu\right] 
\nn\\
&\approx& (\vec{a}\cdot\vec{\zeta})\,\Tr\Big[(-\hat{\nabla}^{-2})^2 \omega_{ab}\,^{\mu}\omega^{ab\nu} \partial_\mu\partial_\nu \Big] + (\vec{b}\cdot\vec{\zeta})\,\Tr\Big[ (-\hat{\nabla}^{-2})^2 \,{\cal R}\,\Delta g^{\mu\nu} \partial_\mu\partial_\nu \Big]\label{xthree}\ee
with $\Delta g^{\mu\nu} = g^{\mu\nu}  - \hat{g}^{\mu\nu}$. In the second line, we have moved  derivatives past some of the fields; the difference only shows up in higher derivative terms in the effective action.

\para
The remaining traces in \eqn{xtwo} and \eqn{xthree} are over momentum. Since we are working on ${\bf R}^3\times {\bf S}^1$, this involves both an integral and a discrete sum\footnote{Strictly speaking, to compute the Wilsonian effective action we should drop the $n=0$ zero-mode in the sum. These terms can be interpreted as counterterms for the 3d theory.} for the momentum $k_4=n/L$, with $n\in {\bf Z}$, for modes on ${\bf S}^1$
\be \Tr \longrightarrow \frac{1}{2\pi L}\sum_{n} \int \frac{d^3 k}{(2\pi)^3}  \nn \ee
With this, the expressions \eqn{xtwo} and \eqn{xthree} for the one-loop contribution to the two-derivative effective action combine to become,
\be S_{\rm 1-loop} = - \frac{1}{2\pi L}\sum_{n} \frac{d^3k}{(2\pi)^3}  &&\ \left\{ (\vec{a}\cdot\vec{\zeta}) \left[\frac{\omega_{abc}\omega^{abc}}{2\hat{g}^{\mu\nu}k_\mu k_\nu} - \frac{\omega_{ab}\,^\mu\omega^{ab\nu} k_\mu k_\nu } {(\hat{g}^{\mu\nu}k_\mu k_\nu)^2} \right] \right. \label{wherewereat}\\ &&\ \ \ \ \ \  +\ (\vec{b}\cdot\vec{\zeta})\left. \left[ \frac {1}{\hat{g}^{\mu\nu}k_\mu k_\nu} 
- \frac{\Delta g^{\mu\nu} k_\mu k_\nu}{(\hat{g}^{\mu\nu}k_\mu k_\nu)^2} \right]   {\cal R} \right\}\nn\ee
 These integrals suffer both quadratic and logarithmic divergences which we need to tame. Our method of choice is Pauli-Villars regularisation.

\subsubsection*{Pauli-Villars Regularisation}

Pauli-Villars offers perhaps the most straightforward method of regularisation. We start by providing all of our original fields with a small mass $m$. This will act as an infra-red cut-off and ultimately we send $m\rightarrow 0$. (In practice, this means that we need only replace $k^2\rightarrow k^2 +m^2$ in the denominator of integrals.)

\para
The UV divergences are tamed by introducing very heavy ghost particles with mass  $\reg$. We will ultimately take $\reg\rightarrow \infty$. Introducing one such field is enough to remove logarithmic divergences, but we also have a quadratic divergence to deal with. This requires the introduction of two further fields; a physical field with mass-squared $\gamma \reg^2$ and a ghost with mass-squared $(\gamma - 1)\reg^2 +m^2$ where $\gamma$ is an arbitrary parameter on which no physical quantity should depend. The upshot is that the integrands in \eqn{wherewereat} are replaced by their regulated form such as 
\be \frac{1}{\hat{g}^{\mu\nu} k_\mu k_\nu} \rightarrow \pv{\frac{1}{\hat{g}^{\mu\nu} k_\mu k_\nu+m^2}} \nn\ee
where we've introduce the notation
\be \pv{f(m^2)} = f(m^2) - f(\reg^2) + f(\gamma \reg^2) - f((\gamma -1)\reg^2 + m^2) \label{pv}\ee
%
%So, for example,
%
%\be \frac{1}{k^2} \rightarrow \pv{\frac{1}{k^2+m^2}} = \frac{1}{k^2 +m^2} -\frac{1}{k^2+M^2} +\frac{1}{k^2 + \gamma M^2} -  \frac{1}{k^2 + (\gamma-1)M^2 +m^2}\nn\ee
%
Our goal is to now evaluate the integrals \eqn{wherewereat} using this  regularisation procedure.

\subsubsection*{Extracting the Divergent Piece}

Before we proceed,  it will help to better understand the origin of the divergent pieces and, more importantly, the finite pieces. 
Because the divergences arise from the UV, it should come as no surprise to learn that they are the same regardless of whether we work on ${\bf R}^4$ or ${\bf R}^3\times {\bf S}^1$. In contrast, the finite terms that we seek are proportional to $1/R^2$ and are only present when we are on the circle. For this reason, it's useful to write 
\be \frac{1}{2\pi L} \sum_{n} \frac{d^3k}{(2\pi)^3} = \int \frac{d^4 k}{(2\pi)^4} + \left[\frac{1}{2\pi L} \sum_{k_4=n/L} \frac{d^3k}{(2\pi)^3} - \int \frac{d^4 k}{(2\pi)^4} \right]\nn\ee
All divergences are contained in the first term. Meanwhile, we will see that the second term, which captures the difference between physics on the circle and in the plane, contains only finite pieces.

\para
As it stands, the integrands in \eqn{wherewereat} are not quite rotationally invariant, even when integrated over ${\bf R}^4$. This is because the background flat metric gives $\hat{g}^{\mu\nu}k_\mu k_\nu = (R^2/L^2) {\bf k}^2 + (L^2/R^2) k_4^2$. To proceed, we rescale the 3-momentum ${\bf k} \rightarrow (R^2/L^2) {\bf k}$. Then, the integrand in \eqn{wherewereat} becomes isotropic.  On grounds of rotational invariance, the divergent piece of the one-loop effective action, arising from integrating over $\int d^4k$, is then given by
\be S_{\rm divergent} &=& - \frac{L^4}{R^2}\int \frac{d^4k}{(2\pi)^4} \ \left\{ (\vec{a}\cdot\vec{\zeta}) \,\left[\frac{1}{2 (k^2+m^2) } - \frac{k^2}{4(k^2+m^2)^2} \right]_{\rm PV} \omega_{abc}\omega^{abc} \right.  \nn\\ &&\ \ \ \ \ \ \ \ \ \ \ \ \ \ \ \ \ \ \ \ \ \ \ \ \left.+\ (\vec{b}\cdot\vec{\zeta}) \left[ \frac {1}{k^2+m^2} - \frac{\hat{g}_{\mu\nu}\Delta g^{\mu\nu} k^2}{4(k^2+m^2)^2} \right]_{\rm PV}   {\cal R}\right\} \label{sdiv}\ee
where the factor of $L^4/R^4$ arises from the aforementioned rescaling of the momentum and is identified as $\sqrt{\hat{g}}$.
 
\para
The regulated integrals in the above expression are easily computed. They are given by
\be \int \frac{d^4k}{(2\pi)^4} \pv {  \frac{1}{k^2 + \mu^2} } = -\frac{1}{16\pi^2}\pv{ m^2 \log m^2}  \nn \ee
%
%where $\sim$ means that we have retained only the terms which remain in the limit $m^2\rightarrow 0$ and $\reg^2\rightarrow \infty$. The other integral gives
and
\be \int \frac{d^4k}{(2\pi)^4} \pv {  \frac{k^2}{\left(k^2 + \mu^2\right)^2} } = -\frac{1}{8\pi^2} \pv{m^2 \log m^2} \nn\ee
We see that the $\omega^2$ terms in \eqn{sdiv} cancel. (This is perhaps rather surprising; if you consider the unregulated integrands with $m^2=0$ then the two terms appear to differ by a factor of 2. But, of course, such unregulated integrals are ill-defined. The same cancelling factor of 2 can also be seen in dimensional regularisation as discussed, for example, in \cite{urbanwarrior}.)

\para
The term proportional to ${\cal R}$ in \eqn{sdiv} does not vanish. Instead, it gives
\be S_{\rm divergent} &=& \frac{\vec{b}\cdot\vec{\zeta}}{16\pi^2}\pv{m^2\log m^2}\left(1-\frac{1}{2}\hat{g}_{\mu\nu}\Delta g^{\mu\nu}\right)\sqrt{\hat{g}}{\cal R} \nn\\  &=& -\frac{15}{64\pi^2}\pv{m^2\log m^2}\sqrt{{g}}{\cal R} \label{newtondiv}\ee
where the $\Delta g$ term acts simply to change the fiducial metric $\sqrt{\hat{g}}$ into the background metric $\sqrt{g}$ (to the order in which we're working). This term is divergent but can be absorbed through a renormalisation of Newton's constant. As we will see in the next section, it agrees with the divergence computed using heat kernel methods. 

\subsubsection*{Extracting the Finite Pieces}

As described above, the finite terms in the effective action \eqn{wherewereat} arise from the difference between physics on the circle and physics on the plane.
\be S_{\rm finite} = -\left[\frac{1}{2\pi L} \sum_{n} \frac{d^3k}{(2\pi)^3} - \int \frac{d^4 k}{(2\pi)^4} \right]  && \left\{ (\vec{a}\cdot\vec{\zeta}) \left[\frac{\omega_{abc}\omega^{abc}}{2\hat{g}^{\mu\nu}k_\mu k_\nu} - \frac{\omega_{ab}\,^\mu\omega^{ab\nu} k_\mu k_\nu } {(\hat{g}^{\mu\nu}k_\mu k_\nu)^2} \right] \right. \nn\\ &&\ \ \ \ \ \  +\ (\vec{b}\cdot\vec{\zeta})\left. \left[ \frac {1}{\hat{g}^{\mu\nu}k_\mu k_\nu} 
- \frac{\Delta g^{\mu\nu} k_\mu k_\nu}{(\hat{g}^{\mu\nu}k_\mu k_\nu)^2} \right]   {\cal R} \right\}\nn\ee
We again rescale the 3-momentum ${\bf k}\rightarrow (R^2/L^2){\bf k}$. Isotropy and parity ensure that the terms with $k_\mu k_\nu$ in the numerator are once again diagonal, but we now have to treat the ${\bf R}^3$ and ${\bf S}^1$ components separately. The relevant integrals are
\be
\frac 1 {2\pi L}\left(\sum_n - \int dn \right) \frac{L^4}{R^4} \int \frac{d^3{\bf k}}{(2\pi)^3} \pv{\frac{1}{(n/L)^2+{\bf k}^2 + m^2}} &\longrightarrow&  \frac{1}{48\pi^2}\frac{L^2}{R^4} \nn \ee
and 
\be 
\frac 1 {2\pi L}\left(\sum_n - \int dn \right) \frac{L^4}{R^4} \int \frac{d^3{\bf k}}{(2\pi)^3} \pv{\frac{{\bf k}^2}{\left((n/L)^2+{\bf k}^2 + m^2\right)^2}} &\longrightarrow& \frac{1}{32\pi^2}\frac{L^2}{R^4} \nn \ee
and
\be
\frac 1 {2\pi L}\left(\sum_n - \int dn \right) \frac{L^4}{R^4} \int \frac{d^3{\bf k}}{(2\pi)^3} \pv{\frac{(n/L)^2}{\left((n/L)^2+{\bf k}^2 + m^2\right)^2}} &\longrightarrow& -\frac{1}{96\pi^2}\frac{L^2}{R^4} \nn
 \ee
where $\longrightarrow$ means that we have dropped terms which vanish as we remove the regulators, $m^2\rightarrow 0$ and $M^2\rightarrow \infty$. This leaves behind only finite contributions as promised. The final result is
\be
S_{\rm finite} = -\int d^4 x  \sqrt g \; \frac{1}{48 \pi^2} \frac{1}{R^2} \Bigg\{(\vec{a}\cdot\vec{\zeta})\, \omega_{ab4}\omega^{ab4} + (\vec{b}\cdot\vec{\zeta})\,{\cal R} \Bigg\}\nn
\ee
 where, as in the divergent case, the role of the $\Delta g$ terms is to ensure that the $R$ that appears here is now the dynamical field rather than the fixed, asymptotic value of $\hat{g}$. Substituting the three-dimensional expressions for $\omega$ and ${\cal R}$ we have
\be
S_{\rm finite} = -\int d^3 x  \sqrt {g_{(3)}} &&\!\! \frac{1}{24\pi} \frac{L}{R^2} \left\{ (\vec{a}\cdot\vec{\zeta}) \, \left[ 2\left( \frac {\partial R}{R}\right)^{2} + \frac{1}{4}\frac{R^4}{L^4}F^{2} \right] \right.\nn \\
 &&\ \ \ \ \ \ \ \ \ +\left.\ (\vec{b}\cdot\vec{\zeta}) \, \left[ {\cal R}_{(3)}+2\left( \frac {\partial R}{R}\right)^{2}+\frac{1}{4}\frac{R^4}{L^4}F^{2}-2\nabla^{2}\log R \right]  \right\}
\nn\ee
We now integrate the last term by parts, discarding the total derivative, leaving us with
\be
S_{\rm finite} = -\int d^3 x  \sqrt {g_{(3)}} \ \frac{1}{24\pi} \frac{L}{R^2} \Bigg\{ (\vec{b}\cdot\vec{\zeta}) \,{\cal R}_{(3)} + 2(\vec{a}-\vec{b})\cdot \vec{\zeta}\left( \frac {\partial R}{R}\right)^{2}  + \frac{1}{4}(\vec{a}+\vec{b} )\cdot\vec{\zeta} \, \frac{R^4}{L^4}F^{2} \Bigg\}\nn
\ee
Note that the finite renormalisations to the scalar $R$ and field strength $F$ are different:. this will prove important shortly since it can be interpreted as a one-loop correction to the complex structure. Putting this together with the tree-level contributions, we find that the one-loop effective action in Euclidean space is given by
\be S_{\rm eff}  &=& \int d^3x\sqrt{g_{(3)}}\,\left\{\frac{1}{2}\left(M_3  + \frac{5}{16\pi}\frac{L}{R^2}\right)\,{\cal R}_{(3)} + \left(M_3 - \frac{1}{6\pi}\frac{L}{R^2}\right)\left(\frac{\partial R}{R}\right)^2\right. \nn\\ &&\ \ \ \ \ \ \ \ \ \ \ \ \ \ \ \ \ \ \ \ \ \ \ \ \  \ \ \ \ \ \ \ \left.+\ \frac{1}{2}\left(M_3 + \frac{11}{24\pi}\frac{L}{R^2}\right) \frac{1}{4}\frac{R^4}{L^4}F^2\right\}
\nn\ee
It remains only to rotate back to Lorentzian signature and to subsequently dualise the photon in favour of the periodic scalar $\sigma$. The result is the  effective action,
\be  {\cal S}_{\rm eff}  &=& \int d^3x\sqrt{-g_{(3)}}\,\left\{\frac{1}{2}\left(M_3  + \frac{5}{16\pi}\frac{L}{R^2}\right)\,{\cal R}_{(3)} - \left(M_3 - \frac{1}{6\pi}\frac{L}{R^2}\right)\left(\frac{\partial R}{R}\right)^2 \right.\nn\\ &&\ \ \ \ \ \ \ \ \ \ \ \ \ \ \ \ \ \ \ \ \ \ \ \ \  \ \ \ \ \ \ \ \left.- \left(M_3 + \frac{11}{24\pi}\frac{L}{R^2}\right)^{-1} \frac{L^2}{R^4}\left(\frac{\partial \sigma}{2\pi}\right)^2
\right\}\label{oneloopact}\ee
as previously advertised in \eqn{loopact}.

\subsection{Supersymmetry and the Complex Structure}\label{susysec}

We now describe how the low-energy effective action is consistent with supersymmetry. After dimensional reduction, the propagating bosonic fields $R$ and $\sigma$ lie in a chiral multiplet \cite{grimm}. (The most general form of the 3d supergravity action with chiral multiplets was presented in \cite{dewitt}.) The lowest component of the chiral multiplet is given by
\be {\cal S} = 2\pi^2 M_{\rm pl}^2 R^2 + i\sigma\label{complexclas}\ee
and the classical action \eqn{clasact} for this complex scalar takes the form
\be S = - M_3\int d^3x\ \sqrt{-g_{(3)}}\  \frac{1}{({\cal S}+ {\cal S}^\dagger)^2}\partial {\cal S}\partial {\cal S}^\dagger\label{sact}\ee
which is derived from  the classical K\"ahler potential
\be K = - \log({\cal S} + {\cal S}^\dagger)\label{kahler}\ee
The presence of the Planck mass $M_{\rm pl}$ in the complex structure \eqn{complexclas} means that this chiral multiplet does not survive the rigid limit in which gravity is decoupled. (The distinction between rigid and gravitational theories was stressed, in particular, in \cite{seiberg}.) This, in turn, means that we cannot use the fact that $R$ sits in a chiral multiplet to restrict the way it appears in superpotentials when rigid supersymmetric gauge theories are compactified on a circle as in \cite{sw3,ahiss}\footnote{We thank N. Seiberg for discussions on these issues.}.

\subsubsection*{One-Loop Corrected Complex Structure}

As we have just seen, the kinetic terms are corrected at one-loop. This in principle affects both the complex structure and K\"ahler potential. For our present purposes, we are only concerned with the shift to the complex structure. 

\para
The renormalisation of the complex structure can be seen from the fact that the $(\partial R)^2$ and $(\partial \sigma)^2$ terms pick up different $1/R^2$ corrections in \eqn{oneloopact}. (Strictly speaking, we should first perform a conformal transformation so that we are working in the Einstein frame, but this only affects the complex structure at order $1/R^4$ and so can be neglected at one-loop order.) It is simple to check that the  one-loop corrected complex structure is given by
\be {\cal S} = 2\pi^2 M_{\rm pl}^2 R^2 + \frac{7}{48}\log(M^2_{\rm pl}R^2) +i\sigma\label{ella}\ee
(Tracing the origin of this shift, we see that it depends on the $\vec{a}$ coefficients defined in \eqn{ttstar}, but is independent of the $\vec{b}$ coefficients defined in \eqn{ebs}.) We will have use for this later when we compute the instanton-generated superpotential.

\subsection{Divergences and the Heat Kernel}\label{heatkernelsec}

The gradient expansion employed in Section \ref{gradsec} is the simplest approach for computing the effective action at the two derivative level. However, it becomes increasingly cumbersome as we look to higher derivatives. In particular, as described at the beginning of Section 3, we are interested in computing the logarithmic running of the coefficient of the Gauss-Bonnet term. For this, we turn to the heat kernel method. The results of this section are not new but, for completeness, we describe the essence of the computation. Further details can be found in the original paper \cite{cduff}. A clear review of heat kernel methods can be found in \cite{heatk}. 

\para
The heat kernel approach starts by writing the one-loop effective action \eqn{ratio} as 
\be S_{\rm 1-loop} = -\sum_{s=0}^2 \zeta_s\,\log\det\Delta_s = \sum_{s=0}^2\zeta_s\int\frac{dt}{t}\ \Tr\,\pv{e^{-t(\Delta_s+m^2)}}\nn\ee
which is true up to an (infinite) constant which we can safely ignore. 
Ultra-violet divergences show up in the $t\rightarrow 0^+$ limit of the integral. The standard expansion gives\footnote{On manifolds with boundary, further terms may arise in the heat kernel approach. These can give rise, for example, to renormalisation of the coefficient of the Gibbons-Hawking term. Here we focus only on bulk divergences.}
\be {\rm Tr} \left[ e^{-t \Delta_s} \right] \sim t^{-2} B_0 + t^{-1} B_2 + B_4 + O(t) \nn \ee
where the Schwinger-DeWitt coefficients $B_k$ are geometric quantities, constructed from the data in the operator $\Delta_s = -\nabla^2 - E_s$, with $\nabla_\mu = \partial_\mu +\frac{1}{2}\omega_{ab\mu}t^{ab}_{(s)}$. The leading divergence  is simply the cosmological constant term,
\be
B_0(\Delta_s) =  \frac{1}{16\pi^2} \int d^4 x \sqrt g \; {\rm tr}\, 1 \nn\ee
This vanishes when we sum over the spins $s=0,1/2,1,3/2,2$ by virtue of supersymmetry, in the guise of $\vec{\zeta}\cdot\vec{d}=0$ as we saw in the previous section. The quadratic divergences are contained in the $B_2$ coefficient which is given by
\be B_2(\Delta_s) =  \frac{1}{16\pi^2} \int d^4 x \sqrt g \; {\rm tr}\, \left( E_s + \frac{1}{6} {\cal R} \right ) \nn \ee
Here, the ${\cal R}/6$ term contains no spin dependence and once again cancels due to supersymmetry. The trace of $E_s$ is given in \eqn{ebs}, leaving us with
\be \sum_s \zeta_s\, B_2(\Delta_s) = -\frac{\vec{b}\cdot\vec{\zeta}}{16\pi^2}\int d^4x\sqrt{g}\ {\cal R}\nn\ee
This is the renormalisation of Newton's constant. One can easily check that it agrees with the quadratic divergence \eqn{newtondiv} that we computed using the gradient expansion in the previous section.

\para
For our purposes, the most important quantities are the logarithmic divergences contained in $B_4$. This is given by
\be
B_4(\Delta_s) & = & \frac{1}{16\pi^2} \int d^4 x \sqrt g \  {\rm tr} \, \Bigg( \frac{1}{6} \nabla^2 E_s  + \frac{1}{6} {\cal R}E_s + \frac{1}{2} E_s^2 + \frac{1}{72} {\cal R}^2 - \frac{1}{180} {\cal R}_{\mu\nu} {\cal R}^{\mu\nu} \ \ \ \ \ \nn \\
 & & \ \ \ \ \ \ \ \ \ \  \ \ \ \ \ \ \ \ \ \ \ \ \ \ \ \ \ \ \ \ \ +\ \frac{1}{180} {\cal R}_{\mu\nu\rho\sigma}{\cal R}^{\mu\nu\rho\sigma} + \frac{1}{48} t^{ab}_{(s)}{\cal R}_{ab\mu\nu}t^{cd}_{(s)}{\cal R}_{cd}^{\ \ \mu\nu}\Bigg ) 
\ \ \ \ \ \ \  \ \ \label{B4}\ee
The story is, by now, familiar. Any terms without spin dependence vanish due to supersymmetry.  The $\nabla^2 E_s$ term survives, but results in divergences for $\nabla^2{\cal R}$ which is a total derivative and vanishes on the backgrounds we're interested in. For this reason, we ignore this term.  Meanwhile, the ${\cal R}E_s$ term results in a logarithmic divergence to ${\cal R}^2$. Both ${\cal R}^2$ terms and ${\cal R}_{\mu\nu}{\cal R}^{\mu\nu}$ terms can be absorbed into the Einstein-Hilbert action through a field redefinition \cite{tv}. Indeed, this is the heart of the statement that the S-matrix of pure Einstein-Hilbert gravity is one-loop finite. 

\para
The upshot of this is that  the only terms that we care about are those that give rise to logarithmic divergences for ${\cal R}_{\mu\nu\rho\sigma}{\cal R}^{\mu\nu\rho\sigma}$. This receives contributions from $E_s^2$ and the last, $t{\cal R}t{\cal R}$ term. In particular, 
\be
{\rm tr} ( E_s^2 ) = c_s {\cal R}_{\mu\nu\rho\sigma}{\cal R}^{\mu\nu\rho\sigma} + \ldots \ \ \ \ \ \ {\rm with}\ \ \ \ \ c_s = \left(0,0,0,2,3\right) \nn\ee
Putting this together with \eqn{ttstar}, we have
\be \sum_{s=0}^2 \zeta_s\, B_4(\Delta_s) = \frac{1}{32\pi^2}\int d^4x\ \sqrt{g} \left(\vec{c}-\frac{1}{12}\vec{a}\right)\cdot\vec{\zeta}\,\left({\cal R}_{\mu\nu\rho\sigma}{\cal R}^{\mu\nu\rho\sigma}+\ldots\right)\nn\ee
The same field redefinitions of the metric that we described above allow us to massage the $\ldots$ terms above so that they become the Gauss-Bonnet term, with the integral given by the Euler character
\be
\chi =\frac{1}{32 \pi ^2} \int d^4 x \sqrt g \; \left( {\cal R}_{\mu\nu\rho\sigma}{\cal R}^{\mu\nu\rho\sigma} - 4 {\cal R}_{\mu\nu}{\cal R}^{\mu\nu} + {\cal R}^2 \right)
\nn\ee
The one-loop effective action therefore contains the logarithmically divergent term
\be S_{\rm one-loop} =-\frac{41}{48} \log(\mu^2/m^2)\,\chi\label{overthecounter}\ee
where, in the Pauli-Villars scheme \eqn{pv}, $\mu^2 = \frac{\gamma-1}{\gamma} \reg^2$. This is the origin of the running of the Gauss-Bonnet coefficient described in \eqn{running}.

\para
We note that the interpretation of this ``running" as a scale-dependent coupling constant comes with a caveat. In gauge theories, the running coupling $g^2(\mu)$ tells us how the strength of local interactions varies with the energy scale of the process. But, in the gravitational context, there is no local process associated to the Gauss-Bonnet term. Instead, it knows only about the global properties of the space. The real physics in this running coupling is the emergence of the infra-red scale $\Lambda_{\rm grav}$ defined in \eqn{lambda} which tells us characteristic scale at which manifolds with different topologies contribute to the path integral.

\section{Non-Perturbative Aspects}\label{instsec}

In this section we describe the instanton corrections to the low-energy effective action. We will show that they generate a superpotential term for the chiral multiplet ${\cal S}$. The techniques of gravitational instanton computations were pioneered in the late 1970s \cite{gravins,thepope, gary} and much of this section is devoted to reviewing and extending this machinery. We start, however, with a brief introduction to gravitational instantons and the role they play in ${\cal N}=1$ supergravity.

\subsection{Gravitational Instantons}\label{gravinssec}

Gravitational instantons are saddle points of the four-dimensional path integral. In supersymmetric theories, we can restrict attention to (anti)-self-dual solutions to the Einstein equations satisfying
\be {\cal R}_{\mu\nu\rho\sigma}  = \pm {}^\star {\cal R}_{\mu\nu\rho\sigma}\label{sd}\ee
Such backgrounds preserve half of the supersymmetry. This means that supersymmetry transformations generate only two fermionic Goldstino zero modes, which is the right number to contribute towards a superpotential in ${\cal N}=1$ theories \cite{wsusy}.  The self-duality requirement \eqn{sd} is a necessary, but not sufficient, condition for instantons to contribute to the superpotential; there may also be further fermionic zero modes which do not arise from broken supersymmetry which we describe below.

\para
For theories on ${\bf R}^3\times {\bf S}^1$, the gravitational instantons are Kaluza-Klein monopoles \cite{sorkin,gp} which, in the present context are perhaps best referred to as ``Kaluza-Klein instantons". From the low-energy 3d perspective, these solutions look like Dirac monopoles and the calculation can be thought of as a gravitational completion of Polyakov's famous computation \cite{polyakov}.  The contribution of these ``Kaluza-Klein instantons" has been discussed previously in the non-supersymmetric context in  \cite{gross} and, more recently, in \cite{sean}. 
\para
The simplest class of gravitational instantons are the multi-Taub-NUT metrics \cite{gravins},
\be 
ds^2 = U({\bf x})d{\bf x}\cdot d{\bf x} + U({\bf x})^{-1}\left(dz + {\bf A}\cdot d{\bf x}\right)^2\label{tn}\ee
with
\be U({\bf x})  = 1 + \frac{L}{2}\sum_{a=1}^k\frac{1}{|{\bf x}-{\bf X}_a|} \ \ \ {\rm and}\ \ \ \nabla\times {\bf A} = \pm \nabla U\nn\ee
The metric is smooth when $z\in [0,2\pi L)$ and the ${\bf X}_a$ are distinct.  For $\nabla\times {\bf A} = \pm \nabla U$, the Riemann tensor obeys ${\cal R}_{\mu\nu\rho\sigma}  = \mp {}^\star {\cal R}_{\mu\nu\rho\sigma}$. 

\para
The Taub-NUT metric takes 
the same form as our Kaluza-Klein ansatz \eqn{kkansatz} with $U = L^2/R^2$. However, because $U\rightarrow 1$ asymptotically, it means that we have made a coordinate choice in which the fiducial length $L$ is taken to be the physical asymptotic length of the circle: $R({\bf x})\rightarrow L$.

\para
One might wonder about the relevance of Taub-NUT spaces to the Euclidean path integral. Ultimately, we're  interested in physics on ${\bf R}^{1,2}\times {\bf S}^1$ and, after a Wick rotation, the boundary of  space  is ${\bf S}^2\times {\bf S}^1$. 
But for  $k\neq 0$, the boundary of the manifold is  the ${\bf S}^1$ is fibered non-trivially over the ${\bf S}^2$. For example, with $k=1$, the boundary is topologically ${\bf S}^3$. The question at hand is whether we should sum over these different boundary conditions in the path integral.
%\footnote{Much of the earlier work on Euclidean quantum gravity focussed on thermodynamic situations in which Euclidean time is compactified. In this situation, it appears that one should {\it not} sum over different windings of  the asymptotic circle.}. 

\para
A similar question arises in gauge theories in flat space, where the issue becomes whether one should sum over topologically non-trivial bundles at infinity. Here the answer is certainly yes: a trivial gauge bundle can be smoothly deformed into an instanton-anti-instanton pair which are subsequently moved far apart. Such configurations certainly contribute to the path integral but  locality and cluster decomposition then requires us to also sum over individual instanton bundles. (See, for example, \cite{seiberg} for a recent discussion of this topic.) However, these same arguments also hold in the present case: we can equally well locally nucleate a NUT-anti-NUT pair which can then be moved far apart. This suggests that should sum over all asymptotic windings. (There is, admittedly, one loophole which is the lack of local observables in a theory of gravity but this does not seem to be a serious objection to the argument.)

\para
Another way to motivate including non-trivial ${\bf S}^1$ bundles is to consider a parallel to a
more familiar story with gauge theory instantons. There, one imposes `initial' and `final'
conditions in Euclidean time and boundary conditions at spatial infinity which require
local decay everywhere, but allow for non-trivial global behaviour of the solution. For us,
where the distinction between initial and boundary conditions is blurred, the obvious
analogy is to consider `initial' and `final' surfaces which are asymptotically flat
hemispheres of ${\bf S}^2$ with a (necessarily trivial) ${\bf S}^1$ bundle, and require them to be glued
in a locally smooth, flat manner. The non-trivial global behaviour now arises due to the
possibility of creating a non-trivial bundle of ${\bf S}^1$ over the whole ${\bf S}^2$.

\para
We conclude that, despite the different boundary conditions, we should be summing over Taub-NUT configurations to determine the low-energy physics on ${\bf  R}^{1,2}\times {\bf S}^1$. We would reach the same conclusion by considering the low-energy world where we would expect to sum over different Dirac monopole configurations provided they have a suitable microscopic completion \cite{polyakov}. We also reach the same conclusion by considering the very high-energy world of string theory, where these Taub-NUT instantons can be viewed as D6-brane instantons wrapping manifolds of G2-holonomy \cite{hmoore}.

\para
The multi-Taub-NUT solution \eqn{tn} enjoys $3k$ bosonic zero modes, parameterised by the centres ${\bf X}_a$,  and $2k$ spin-3/2 fermionic zero modes \cite{thepope}\footnote{For Yang-Mills instantons, the number of zero modes can be simply determined by integrating the anomaly. In the present case there is a mismatch between the integrated anomaly \eqn{anomaly} and the number of zero modes due to the presence of boundary terms. These are known as eta-invariants \cite{egh} and will also play a role when we come to discuss the one-loop determinants around the background of the gravitational instanton.}. Although this result is well known, we will provide a slightly different derivation of the index theorem for the fermionic zero  modes  in Section \ref{jacsec} en route to calculating the one-loop determinants. For now, we merely note that only the $k=1$ Taub-NUT solution, with two fermionic zero modes, can contribute to the superpotential \cite{grimm}.

\subsubsection*{The Action}

The Einstein-Hilbert action evaluated on the Taub-NUT  space with charge $k=1$ is, after subtracting appropriate counterterms,  given by \cite{gary,clifford}, 
\be S_{\rm TN} = 2\pi^2 M_{\rm pl}^2 R^2 \nn\ee
where $R$ here is interpreted as the asymptotic radius of the circle. (In the coordinates \eqn{tn}, we could just as well have written $S_{\rm TN} = 2\pi^2 M_{\rm pl}^2 L^2$.) However, there are a number of further contributions to the action. The first comes from the dual 3d photon which, as first observed by Polyakov, acts as a chemical potential for the topological instanton charge \cite{polyakov}. This follows from the coupling \eqn{sigmacoupling}: the 3d field strength arising from the metric \eqn{tn} has charge $\int_{\rm S^2} F = 2\pi L$, which ensures that the single Taub-NUT instanton also comes with a factor of 
\be {\cal S} = 2\pi^2 M_{\rm pl}^2 R^2 + i\sigma \nn\ee
This coincides with the classical complex structure \eqn{complexclas}. Of course, this had to be the case since the superpotential will come with the factor ${\cal W}\sim e^{-{\cal S}}$. Turning this observation on its head, it could be viewed   as a particularly simple derivation of the action of Taub-NUT, a subject which has previously enjoyed some controversy before the definitive analysis of \cite{clifford}.

\para
Further contributions come from the total derivative terms: these are the Gauss-Bonnet term \eqn{gb} and the Pontryagin term \eqn{pont}. 
For Taub-NUT, the integral of the Gauss-Bonnet term gives the Euler character (there is no boundary contribution),  
\be \chi = \frac{1}{32\pi^2} \int d^4x\sqrt{g}  \ {}^\star{\cal R}^\star_{\mu\nu\rho\sigma}\,{\cal R}^{\mu\nu\rho\sigma} = 1\nn\ee
This means that   the Taub-NUT instanton will contribute to the superpotential in the form
\be {\cal W}\sim  e^{-{\cal S}}e^{-\tau_{\rm grav}}\label{wtn}\ee
This is the promised superpotential \eqn{w}. Here $\tau_{\rm grav}$ is given by \eqn{taugrav} and, like ${\cal S}$, is naturally complex and lives is the lowest component of a chiral multiplet.  This superpotential drives the moduli ${\cal S}$ to large values, decompactifying the Kaluza-Klein circle.

\subsubsection*{A Summary of What's to Come}

 The rest of this section is devoted to understanding more fully the computations involved in deriving \eqn{wtn}.  The key extra ingredient is the computation of the one-loop determinants around the background of  Taub-NUT. We will show that, despite the existence of supersymmetry, these determinants do not cancel. Instead, after removing the zero-modes, the determinants are computed to be (up to a numerical constant)
 \be {\rm dets}  \sim \mu^{41/24} R^{-7/24}\nn\ee
 where $\mu$ is the UV cut-off. This provides the prefactor to the superpotential \eqn{wtn} which  becomes
 \be {\cal W} \sim \mu^{41/24} R^{-7/24} e^{-{\cal S}} e^{-\tau_{\rm grav}}\nn\ee
 Now we can see how all the pieces fit together. As we explained in Section \ref{pertsec}, the Gauss-Bonnet coupling $\tau_{\rm grav}$ runs at one-loop and so depends on $\mu$. This combines with the $\mu^{41/24}$ factor that arises from the determinant and  whose exponent agrees with the beta-function for $\tau_{\rm grav}$. Together they  give  the RG-invariant scale $\Lambda_{\rm grav}$ defined in \eqn{lambda}. Meanwhile, the factor of $R^{-7/24}$ coming from the determinants can be exponentiated and has the right coefficient to shift the chiral multiplet ${\cal S}$ to its one-loop corrected value given in \eqn{ella}. The net result is that the superpotential takes the simple form
 \be {\cal W} \sim \Lambda_{\rm grav}^{41/24}\,e^{-{\cal S}}\nn\ee
 where the $\sim$ is hiding a numerical coefficient and factors of $M_{\rm pl}$ which ensure that the dimensions work out.
 
 \para
 The remainder of this section is devoted to performing these computations in some detail. However,  before diving into this, we make a few more general comments on these instanton computations. 
 
 \subsubsection*{Relation to Three Dimensional Gauge Theories}

There is a close analogy between our gravitational computation and the quantum dynamics of ${\cal N}=1$   $SU(2)$ Yang-Mills theory compactified on ${\bf R}^{1,2}\times {\bf S}^1$. In both cases, the low-energy physics comprises of a $U(1)$ gauge field and a neutral scalar, with the only difference classically lying in the form of the K\"ahler potential \eqn{kahler}. 

\para
In the case of Yang-Mills theory, there are two contributions to the low-energy effective action. The first, considered long ago in \cite{ahw}, arises from monopoles in the three-dimensional effective gauge theory and results in a run-away potential on the Coulomb branch, parameterised by the chiral multiplet $\Phi$. The second contribution is four-dimensional in origin; it arises from monopoles twisted around the spatial ${\bf S}^1$, sometimes known as calarons. This second contribution carries the quantum numbers of a four-dimensional instanton, $e^{2\pi i\tau_{YM}}$ with $\tau_{YM} = 2\pi/\theta_{YM} + 4\pi i /g^2_{YM}$. The net result is the superpotential \cite{sw3,ahiss},
\be {\cal W}_{YM} \sim e^{-\Phi} + e^{+\Phi}e^{2\pi i\tau_{\rm YM}}\nn\ee
The gravitational instanton contribution \eqn{wtn} is analogous to the second term above\footnote{We thank N. Seiberg for discussions on this issue.}. Both are associated to physics in four dimensions that does not have a strict three-dimensional counterpart. And both drive the moduli to the region where the heavy states -- whether W-bosons or Kaluza-Klein modes -- become light. In the Yang-Mills case, this is the strongly coupled region and the W-bosons do not ultimately become massless; in the gravitational case, this is the weakly coupled region and the Kaluza-Klein modes do become massless.

\para
Of course, in the Yang-Mills case the first term stabilises the Coulomb branch scalar and the theory on ${\bf S}^1$ has two, isolated vacua. There seems to be no analog of the first term in the gravitational context. The reason is simply that the strict three dimensional theory is $U(1)$ and not $SU(2)$ and the former has no microscopic monopoles of its own.

\subsubsection{Other Topologies and Moduli Fixing}

The Taub-NUT metrics \eqn{tn} are not the only self-dual gravitational instantons which asymptote to a space with one compact direction. For our purposes, the other relevant instanton is the Atiyah-Hitchin manifold ${\cal M}_{AH}$. This admits a smooth hyperK\"ahler metric with isometry group $SO(3)$, as opposed to the $SO(3)\times U(1)$ isometry of Taub-NUT \cite{ah}. This  means that the Kaluza-Klein modes around the asymptotic ${\bf S}^1$ are excited in this solution.

\para
The Atiyah-Hitchin manifold has 3 bosonic zero modes and 2 fermionic zero modes, the right number to contribute to the superpotential\footnote{The double cover of Atiyah-Hitchin also admits a smooth hyperK\"ahler metric, but this space has 6 bosonic zero modes and 4 fermionic zero modes so cannot contribute to the superpotential. (The 6 bosonic zero modes consist of 3 translations and 3 deformations described in \cite{dancer}.)}. Let us first proceed naively and ask what would happen if we {\it were} to admit Atiyah-Hitchin as a contribution to the path integral. While Taub-NUT has winding, or magnetic charge, 1, Atiyah-Hitchin has winding number -4. (See, for example, \cite{sw3}.) By supersymmetry, this means that its complexified action should be
$S_{AH} = -4{\cal S}$. The minus sign is important here. It is related to the fact that, viewed as a soliton, Atiyah-Hitchin has negative mass. As explained in \cite{sean}, it follows from the breaking of the $U(1)$ isometry, and the fact that spatial kinetic terms act like a negative mass in gravity. 
(It is also related to the fact that M-theory compactified on Atiyah-Hitchin reduces to type IIA string theory in the presence of an orientifold $O6$-plane \cite{seib,sen} and orientifolds have negative tension.)  Including contributions from both Taub-NUT and Atiyah-Hitchin would give rise to the superpotential,
\be {\cal W} \sim e^{-{\cal S}}e^{ - \tau_{\rm grav}} + e^{+4{\cal S}}e^{-\tau_{\rm grav}}\nn\ee
The theory appears to now have a ground state with the radius  $R$ fixed at some value (albeit at the Planck scale where the analysis is not trustworthy). The presence of the Atiyah-Hitchin manifold here is reminiscent of the role orientifolds play in more complicated models of moduli stabilisation \cite{kklt}. 

\para
Nonetheless, there is reason to doubt that we should include ${\cal M}_{AH}$ as a saddle in the path integral. 
This is because the asymptotic structure of ${\cal M}_{AH}$  is given by a ${\bf S}^1$ bundle over ${\bf RP^2}\cong{\bf S}^2/{\bf Z}_2$ rather than a bundle over ${\bf S}^2$. It's not clear whether such an asymptotic change of topology should be allowed in the sum over geometries.

\para
Of course, we have just argued that we should be summing over different asymptotic ${\bf S}^1$ bundles and we could try to repeat the nucleation argument that we made above for NUTs. Now the object that lies at the centre of Atiyah-Hitchin is a ``bolt", a 2-cycle with topology ${\bf RP}^2$ and size $\sim R$. The non-local nature makes it less clear whether bolts and anti-bolts  can be smoothly nucleated from the vacuum. Furthermore, the ``gluing" argument that we presented above suggests that we should not include Atiyah-Hitchin in the path integral.

\para
While we don't yet know the complete rules for performing the path integral over manifolds we 
different topology, we suspect that it is consistent to only include ${\bf S}^1$ bundles over ${\bf S}^2$ in the path integral. This means that we do not sum over discrete quotients of the asymptotic ${\bf S}^2$ and ignore the contribution from Atiyah-Hitchin. The same conclusion was reached in \cite{sean}. In the remainder of the paper, we proceed under this assumption.  
%Clearly it would be useful to have a cleaner understanding of these old issues.

\subsection{Determinants Again}\label{instdetsec}

In Section \ref{detsec} we computed the ratio of determinants arising from one-loop fluctuations around a general background. They are
 \be  \Gamma = \frac{\det\Delta_{1} \,\det^{1/4}\Delta_{3/2}}{\det^{1/2}\Delta_{2}\,\det^{1/2}\Delta_0\,\det^{3/4}\Delta_{1/2}}\label{oldratio}\ee
 where $\Delta_s$ is the Laplacian-type operator acting on a field of spin $s$. The definitions of each of them can be found in Section \ref{detsec}. The purpose of this section is to compute this ratio of determinants explicitly in the Taub-NUT background. We will find that, despite the existence of supersymmetry, the bosonic and fermionic determinants do not cancel. Nonetheless, there is sufficient simplification that the ratio can be evaluated exactly.
 
\subsubsection{Determinants in an Anti-Self-Dual Background}\label{asdbacksec}

We start by finding a simplified expression for the ratio of determinants in an anti-self-dual background obeying ${\cal R}_{\mu\nu\rho\sigma}  = - {}^\star {\cal R}_{\mu\nu\rho\sigma}$. The key observation is that the self-dual part of the spin connection is flat. This means that it is possible to choose coordinates such that,
\be \omega_{ab}^{\ \ \mu} = -\frac{1}{2}\epsilon_{abcd}\,\omega^{cd\mu}\label{sdspin}\ee
(One can check that the coordinates in which we've written the Taub-NUT metric \eqn{tn} have this property.)

\para
To see the implications of this, it is useful to change from the Majorana basis of gamma matrices introduced in \eqn{majgamma} to a chiral basis. In Euclidean space, these are given by
\be \gamma^a = \left(\begin{array}{cc} 0 & \sigma^a\\ \bar{\sigma}^a & 0 \end{array}\right)\ \ \ \ \ \ a=1,2,3,4\ \ \ \ {\rm and}\ \ \ \  \gamma^5 = \left(\begin{array}{cc} 1& 0\\  0 & -1 \end{array}\right)\nn\ee
with $\sigma^a = (1,\vec{\sigma})$ and $\bar{\sigma}^a = (1,-\vec{\sigma})$. In such a basis, Dirac spinors decompose in the familiar left-handed (undotted) and right-handed (dotted) chiral spinors,
\be \psi = \left(\begin{array}{c} \chi_\alpha \\ \bar{\lambda}^{\dot{\alpha}}\end{array}\right)\label{psidecomp}\ee
In what follows, we work with this chiral decomposition, using indices $\alpha,\dot{\alpha}=1,2$. (This contrasts with the earlier part of the paper where we worked with 4d Majorana spinors.) The utility of this is that the spin connection acting on right-handed  spinors is $\frac{1}{2} \omega_{ab\mu} \bar{\sigma}^{ab}_{\dot{\alpha}\dot{\beta}}$ where $\bar{\sigma}^{ab} = \frac{1}{2}\bar{\sigma}^{[a}\sigma^{b]}$ is self-dual and so, in the coordinates in which \eqn{sdspin} holds, vanishes when contracted with the spin connection. Meanwhile, the spin connection acting on left-handed spinors involves $\omega_{ab\mu}\sigma^{ab}_{\alpha\beta}$ and does not vanish since $\sigma^{ab} = \frac{1}{2}\sigma^{[a}\bar{\sigma}^{b]}$ is anti-self-dual. This means that the chiral Dirac operator acting on left-handed fermions  -- which we call  $\bar{\sigma}^\mu\nabla_\mu^+$ -- includes a  spin connection, but  the chiral operator acting on right-handed fermions --  which we call $\sigma^\mu \nabla_\mu^-$ -- does not. (Of course, both of these covariant derivatives do contain Levi-Civita connections when acting on objects which also carry vector indices.)

\para
We will show that each of the operators $\Delta_s$ has a natural decomposition into operators that act  on left-handed or right-handed spinors. This is simplest to see for the spin $s=1/2$ operator, where we have
\be \Delta_{1/2} = (i\Dslash)^2 = \left(\begin{array}{cc} -\sigma^\mu\nabla_\mu^-\bar{\sigma}^\nu\nabla^+_\nu & 0 \\ 0 & -\bar{\sigma}^\mu\nabla_\mu^+{\sigma}^\nu\nabla^-_\nu\end{array}\right) \equiv \left(\begin{array}{cc} \Delta_{1/2\,+} & 0 \\ 0 & \Delta_{1/2\,-}\end{array}\right)\nn\ee
Moreover,  the self-duality of the spin connection means that the operator on right-handed fermions simplifies yet further. It is given by 
\be \Delta_{1/2\,-} = \Delta_0\,{\bf 1}_2\ \ \ \ \Rightarrow \ \ \ \ \det\Delta_{1/2\,-} = (\det\Delta_0)^2\nn\ee
To perform a similar decomposition for higher spin operators, we need to work a little harder. We start with $\Delta_1$ defined in \eqn{delone}. A self-dual background has ${\cal R}_{\mu\nu} =0$, so the operator involves only the Laplacian acting on vectors. To decompose this in terms of spinors, we use the fact that the background admits two, orthogonal, covariantly constant (and, in fact, actually constant) right-handed spinors, $\xi^{\dot{\alpha}}_{(i)}$. These obey the simple equation
\be \nabla_\mu^- \xi_{(i)} = 0\ \ \ \ \ i=1,2\nn\ee
where, in the coordinates in which \eqn{sdspin} holds,  $\nabla_\mu^-\xi_{(i)} \equiv \partial_\mu\xi_{(i)}$. 

\para
The constant spinors $\xi_{(i)}$ allow us to decompose any (complexified) field so that the dynamical degrees of freedom live in irreducible representations of $su(2)_L\subset so(4)$.
We first demonstrate this with the vector field $A^a$ which we write it in the usual bi-spinor form as $A^{\alpha\dot{\alpha}} = A_a(\sigma^a)^{\alpha\dot{\alpha}}$. The existence of the pair of constant spinors $\xi^{\dot{\alpha}}$ allows us to write a general complex vector in this background as
\be A^{\alpha\dot{\alpha}} = \sum_{i=1}^2 \,a^\alpha_{(i)}\xi^{\dot{\alpha}}_{(i)}\nn\ee
where the dynamical degrees of freedom are now the two left-handed spinors $a_{(i)}$. When sandwiched between two such vectors, $\Delta_1$ reads
\be
\tilde{A}^{\dagger\, a}(\Delta_1)_{a;b} A^b = \sum_{i=1}^2 \xi^{\dagger}_{(i)}\xi_{(i)} \, (\tilde{a}^\dagger_{(i)}\nabla^2a_{(i)})\nn\ee
where we've used the fact that the $\xi_{(i)}$ with $i=1,2$ are orthogonal to eliminate the cross-terms. The upshot of this argument is that in a self-dual background, we can write
\be \det\Delta_1 = (\det\Delta_{1/2\,+})^2\nn\ee

Let's now move on to discuss $\Delta_{3/2}$ defined in \eqn{delthree}. This involves a new element since the Riemann tensor now appears. We make use of the fact that, after replacing the spatial indices with bi-spinors, an anti-self-dual Riemann tensor can be written as,
\be {\cal R}_{\alpha\dot{\alpha}\,\beta\dot{\beta}\,\gamma\dot{\gamma}\,\delta\dot{\delta}} = {\cal C}_{\alpha\beta\gamma\delta} \,\epsilon_{\dot{\alpha}\dot{\beta}}\,\epsilon_{\dot{\gamma}\dot{\delta}} \nn \ee
where ${\cal C}_{\alpha\beta\gamma\delta}$ is the totally symmetric, anti-self-dual Weyl tensor. 
As in the spin-1/2 case, the $\Delta_{3/2}$ operator naturally decomposes into left and right-moving parts, 
\be \det \Delta_{3/2} = \det\Delta_{3/2\,-} \,\det\Delta_{3/2\,+}\nn\ee
To get more of a handle on these determinants, we again decompose a spin-3/2 fermion in terms of the covariantly constant spinors $\xi_{(i)}$. We have to treat the left and right-moving pieces somewhat differently. A general, complex right-handed spinor can be decomposed as 
\be \psi^{\alpha\dot{\alpha}\dot{\beta}}\equiv(\sigma^\mu)^{\alpha\dot{\alpha}}\psi_\mu^{\dot{\beta}}  = 
 f_{(1)}^\alpha \ks{1}{\alpha}\ks{1}{\beta} + f_{(2)}^\alpha \ks{1}{\alpha}\ks{2}{\beta} +f_{(3)}^\alpha \ks{2}{\alpha}\ks{1}{\beta} +f_{(4)}^\alpha \ks{2}{\alpha}\ks{2}{\beta}  \nn\ee
The 8 dynamical degrees of freedom are now contained in four, left-moving spinors $f_{(i)}$, $i=1,2,3,4$. Perhaps unsurprisingly, the Riemann tensor does not act on this part of $\psi$. The same kind of argument that we used for $\Delta_1$ shows that $\Delta_{3/2\,-}$ does not mix the different $f_{(i)}$, and we find
\be \det\Delta_{3/2\,-} = (\det\Delta_{1/2\,+})^4\nn\ee
The decomposition of the left-handed spin-3/2 field involves a new ingredient. We write
\be \psi^{\alpha\dot{\alpha}{\beta}}\equiv(\sigma^\mu)^{\alpha\dot{\alpha}}\psi_\mu^{{\beta}}  = 
 \sum_{i=1}^2\ F_{(i)}^{\alpha\beta} \ks{i}{\alpha} + \phi_{(i)}\epsilon^{\alpha\beta} \ks{i}{\alpha} 
 \label{gd1}\ee
 Now the dynamical degrees of freedom are contained in two scalars $\phi_{(i)}$ and two, symmetric tensors $F_{(i)}^{\alpha\beta}$. The Riemann tensor does not affect the scalar fields $\phi_{(i)}$; these merely contribute a factor of $(\det\Delta_0)^2$ to $\det\Delta_{3/2\,+}$. However, the Riemann tensor does affect the operators acting on the symmetric tensors $F_{(i)}$. To see how, we look at the contraction
 \be
(\bar{\sigma}^a)^{\dot{\alpha}\alpha}\left(\Delta_{3/2\;+}\right)_{a\gamma;b}\,^\delta\,(\sigma^{b})_{\beta\dot{\beta}} &=& \bar{\sigma}^{a\;\dot{\alpha}\alpha}\left( - \eta_{ab}\delta_\gamma^\delta \nabla^2 - \frac 1 2 {\cal R}_{cdab}\sigma^c_{\gamma \dot{\gamma}} {\bar\sigma}^{d\;\dot{\gamma}\delta} \right)\sigma^{b}_{\beta\dot{\beta}} \nn \\
%&=& -2 \delta^{\dot{\alpha}}_{\dot{\beta}} \delta^\alpha_\beta \delta^\delta_\gamma \nabla^2 - \frac 1 2 {\cal R}_{\gamma\dot{\gamma}}\,^{\delta \dot{\gamma}\alpha \dot{\alpha}}\,_{\beta\dot{\beta}} \nn \\
&=& -2 \delta^{\dot{\alpha}}_{\dot{\beta}}\, \delta^\alpha_\beta\, \delta^\delta_\gamma \nabla^2 - \frac 1 2 {\cal C}_{\gamma}\,^{\delta \alpha }\,_{\beta}\, \delta_{\dot{\gamma}}^{\dot{\gamma}}\,\delta^{\dot{\alpha}}_{\dot{\beta}} \nn \\
&\equiv& 2\delta^{\dot{\alpha}}_{\dot{\beta}} \  \Delta_{C}\,^{\alpha}\,_{\gamma;}\,_{\beta}\,^\delta
\nn \ee
where we define a new operator $\Delta_{C}$ which acts on anti-self-dual 2-forms (which transform in the $(1,0)$ representation of $SO(4)$ rotations) and 
involves the Weyl tensor:
\be \left(\Delta_{C}\right)^{\alpha\beta}\,_{\gamma\delta} = -\delta^\alpha_{\gamma}\, \delta^\beta_{\delta} \nabla^2 - \frac{1}{2} C^{\alpha\beta}_{\ \ \gamma\delta} \nn\ee
Our  expression for the left-moving spin-3/2 determinant is then
\be \det\Delta_{3/2\,+} = (\det\Delta_C)^2\,(\det\Delta_0)^2\nn\ee
The same operator $\Delta_C$ also shows up in the determinant of $\Delta_2$. The traceless part of the metric is decomposed as 
\be
\bar{h}^{\alpha\dot{\alpha}\,\beta\dot{\beta}} = H_{(1)}^{\alpha\beta} \ks{1}{\alpha}\ks{1}{\beta} +  H_{(2)}^{\alpha\beta} \xi_{(1)}^{(\dot{\alpha}}\ks{2}{\beta)} +  H_{(3)}^{\alpha\beta} \ks{2}{\alpha}\ks{2}{\beta}
\label{gd2} \ee
where the nine dynamical degrees of freedom are now contained in three symmetric tensors, $H_{(i)}^{\alpha\beta}$, with $i=1,2,3$. The Laplacian operator $\Delta_2$ is defined in \eqn{deltwo} and also contains a Riemann tensor term. To understand its action on the $H_{(i)}$, we again look at the contraction
%
%
%%% Relevant convention here is that $X_{\alpha\dot{\alpha}} = X_a \sigma^a_{\alpha\dot{\alpha}}$
%
\be
\bar{\sigma}^{a\;\dot{\alpha}\alpha} \bar{\sigma}^{b\;\dot{\beta}\beta} (\Delta_{2})_{ab;cd} \sigma^{(c}_{\gamma\dot{\gamma}}\sigma^{d)}_{\delta\dot{\delta}} &=& \bar{\sigma}^{a\;\dot{\alpha}\alpha} \bar{\sigma}^{b\;\dot{\beta}\beta} \left( -\frac{1}{4} \eta_{ac}\eta_{bd}\nabla^2 - \frac 1 2 {\cal R}_{acbd} \right) \sigma^{(c}_{\gamma\dot{\gamma}}\sigma^{d)}_{\delta\dot{\delta}} \nn\\
&=& \frac{1}{2}\left[ -\delta^\alpha_{\gamma} \,\delta^\beta_{\delta}\,\delta^{\dot{\alpha}}_{\dot{\gamma}}\, \delta^{\dot{\beta}}_{\dot{\delta}} \,\nabla^2 - \frac{1}{2} {\cal R}^{\alpha\dot{\alpha}}\,_{\gamma\dot{\gamma}}\,^{\beta\dot{\beta}}\,_{\delta\dot{\delta}} + ({\gamma\dot{\gamma} \leftrightarrow \delta\dot{\delta}) } \right]
\nn\\ &=& \frac{1}{2} \delta^{\dot{\alpha}}_{\dot{\gamma}}\, \delta^{\dot{\beta}}_{\dot{\delta}} \,\left(\Delta_{C}\right)^{\alpha\beta}\,_{\gamma\delta} 
\nn \ee
where, at each step, these operators are understood to be acting on suitably symmetrised objects. This means that we have
\be \det\Delta_2 = (\det\Delta_C)^3 = \frac{(\det\Delta_{3/2\,+})^{3/2}}{(\det\Delta_0)^3}\nn\ee
Putting all this together, we find that the ratio of determinants \eqn{oldratio} in an anti-self-dual background can be written as
\be \Gamma  = \left(\frac{\det \Delta_{3/2\; +}}{\det \Delta_{3/2\; -}} \right)^{-1/2} \left(\frac{\det \Delta_{1/2\; +}}{\det \Delta_{1/2\; -}} \right)^{+1/4}
\label{newratio} \ee
The determinants take the form of ratios of chiral Dirac operators. This is characteristic of instanton computations in supersymmetric theories. Indeed, since the spectrum of non-vanishing eigenvalues of $\Delta_{s\,+}$ (with $s=1/2,3/2$) is identical to the spectrum of $\Delta_{s\,-}$  one might naively think that these determinants cancel. (This was the conclusion reached in \cite{thepope} based on an explicit bijection between the bosonic and fermionic eigenfunctions in of the operators in \eqn{oldratio}.) However, this is too quick. The spectra of both $\Delta_{s\,+}$ and $\Delta_{s\,-}$ contain a continuum of scattering states, and while the range of eigenvalues of the two operators coincide, their densities are not necessarily the same.  Below we will compute $\Gamma$ in a multi-Taub-NUT background and show that it is non-trivial.

\para
The non-cancellation of determinants around self-dual backgrounds has precedent. It occurs in three-dimensional supersymmetric gauge theories where the instantons are 't Hooft-Polyakov monopoles \cite{dmktv,dtv}. (The spectral asymmetry of the Dirac operators had been appreciated earlier in the renormalisation of monopole states in four-dimensional gauge theories \cite{kaul}.) The non-cancellation of determinants  also arises in supersymmetric quantum mechanics where the instantons are kinks \cite{pst}. (Again, the first appearance of this can be traced to the mass renormalisation of kinks in two dimensional theories \cite{dorey}; a detailed review of these effects can be found in \cite{anton}.)

\para
Finally,  we mention that closely related results have been seen recently in the computation of the elliptic genus in non-compact sigma-models, where the non-cancellation of a continuum of scattering states results in a holomorphic anomaly \cite{jan,sameer}. This effect also occurs for Taub-NUT sigma-models \cite{sameer2}. It would be interesting to see if there is any deeper relationship between these two effects.

\subsubsection{Evaluating the Determinants}

We now turn to the task of evaluating the determinants explicitly. This is possible because there is a close relationship between the ratio of determinants in \eqn{newratio} and the (regularised) index for the appropriate Dirac operator \cite{dmktv}. To see this we first define the regularised ratio
\be D(m^2) = \frac{\det\Delta_++m^2}{\det\Delta_-+m^2}\label{dm}\ee
This expression could apply to either $s=1/2$ or $s=3/2$ operators. Here $m^2$ plays the role of an infra-red regulator; its presence will  allow us to easily extract the zero modes from the determinants later. Now consider
\be {\cal I}(m^2)  = \frac{\partial \log D}{\partial \log m^2}  =   {\rm Tr} \left[ \frac {m^2}  {\Delta_+ + m^2 } - \frac {m^2} { \Delta_- + m^2 } \right]\nn\ee
This is the regularised index of the Dirac operator. The index itself is given by
\be {\cal I} = \lim_{m^2\rightarrow 0} {\cal I}(m^2)\nn\ee
and counts $n_+ - n_-$ where  $n_\pm$ is the number of zero modes of $\Delta_\pm$. 

\para
In what follows, we want to treat both $s=1/2$ and $s=3/2$ operators at once. We can do this at the expense of introducing some new notation. We return to the original 4-component spinor notation, with the Dirac operator written as $\hat{\gamma}\cdot \nabla$. For the spin-1/2 field, we simply choose  $\hat{\gamma}^\mu=\gamma^\mu$. But, for the spin-3/2 field, the Dirac operator in \eqn{readyset} means we should pick  $\left(\hat{\gamma}^{\mu}\right)_{\rho\sigma}=-\frac{1}{2}\gamma_{\sigma}\gamma^{\mu}\gamma_{\rho}$, where the additional indices are contracted  with the spacetime indices of $\psi_\mu$. 

\para
For both cases, we have $\{\hat{\gamma}^a,\hat{\gamma}^b\} = 2 \delta^{ab}$, and $\hat{\gamma}^5 = \hat{\gamma}^1\hat{\gamma}^2\hat{\gamma}^3\hat{\gamma}^4 = \gamma^5$ so that $\{\hat{\gamma}^5,\hat{\gamma}^a\} = 0$. We should also bear in mind that the Lorentz generators $t^{ab}$  are different for the two spins.

\para
With this new notation, we can write the regularised index as
\be {\cal I}(m^2)= {\rm Tr}  \left[ \hat{\gamma}^5 \frac {m^2} {-(\hat{\gamma} \cdot \nabla)^2 + m^2 } \right]
\nn\ee
We now split this expression for ${\cal I}(m^2)$ into two terms. One of these will be somewhat subtle and we should be careful in proceeding. Wary of this, we will work with a form of zeta-function regularisation. This means first introducing a new parameter $z$ and replacing the expression in square brackets above with
\be
 \hat{\gamma}^5 \frac {m^2} {\left(-(\hat{\gamma}\cdot\nabla)^2 + m^2\right)^{1+z} } =   \hat{\gamma}^5 \frac {1} {\left(-(\hat{\gamma}\cdot\nabla)^2 + m^2\right)^{z} } + \hat{\gamma}^5 \frac {(\hat{\gamma}\cdot\nabla)^2} {\left(-(\hat{\gamma}\cdot\nabla)^2 + m^2\right)^{1+z} }
\ \ \ \ \ \label{twoterms}\ee
We will ultimately set $z=0$. The first term above naively looks like it reduces to $\hat{\gamma}^5$ when we set $z=0$. But this is too hasty: it ignores the presence of the anomaly. To see this, we use the same heat kernel techniques that we employed in Section \ref{heatkernelsec}. Taking the trace, the first term above reads
\be
{\rm Tr}\left[ \hat{\gamma}^5 \frac {1} {\left(-(\hat{\gamma}\cdot\nabla)^2 + m^2\right)^{z} }\right] = {\rm Tr}\left[ \hat{\gamma}^5 \frac{1}{\Gamma(z)}\int_{0}^{\infty}\frac{\mathrm{d}t}{t^{1-z}}e^{-\left(-\left(\hat{\gamma}\cdot\nabla\right)^{2}+m^{2}\right)t}\right] \nn\ee
This is the same kind of integral that we saw in Section \ref{heatkernelsec}. Up to terms which vanish as $z\rightarrow 0$, the result is very almost the expression $B_4$ given in \eqn{B4}; the only difference is the presence of $\hat{\gamma}^5$ in the spinor trace. This kills most of the terms and changes ${\cal R}_{\mu\nu\rho\sigma}{\cal R}^{\mu\nu\rho\sigma}$ expression in \eqn{B4} into ${}^* {\cal R}_{\mu\nu\rho\sigma}{\cal R}^{\mu\nu\rho\sigma}$. The end result is
\be \lim_{z\rightarrow 0}
{\rm Tr}\left[\hat{\gamma}^5 \frac {1} {\left(-(\hat{\gamma}\cdot\nabla)^2 + m^2\right)^{z} }\right] =\frac{\alpha_s}{24 \cdot 16\pi^2} \int d^4 x \sqrt{g} \  {}^\star {\cal R}_{\mu\nu\rho\sigma}{\cal R}^{\mu\nu\rho\sigma} \nn\ee
This is the promised contribution from the axial anomaly. The coefficient $\alpha_s$ depends on the spin of the operator and is given by\footnote{In the expression for the axial anomaly \eqn{anomaly}, the spin-3/2 and spin-1/2 contributions differ by a factor of $-21$. This is because, in computing the physical anomaly, the factor of $-21$ includes the contribution from three spin-1/2 ghosts. These have different chiral charges and change the $\alpha_{3/2}=-20$ that arises in the present computation into the $-21$ that appears in \eqn{anomaly}.}
\be \alpha_{1/2} = 1\ \ \ {\rm  and}\ \ \  \alpha_{3/2} = -20\nn\ee
We now turn to the second term in \eqn{twoterms}. This term is less delicate and we can happily set $z=0$ from the beginning without repercussion. (We will, however, still implicitly use zeta-function regularisation later when we come to evaluate it.) This term is, in fact, a total derivative, and the full regularised index takes the form
\be
{\cal I}(m^2) =  \frac{\alpha_s}{24 \cdot 16\pi^2} \int d^4 x \sqrt{g} \,^\star {\cal R}_{\mu\nu\rho\sigma}{\cal R}^{\mu\nu\rho\sigma} + \int dS_\mu  \,\sqrt{g_{\rm bdy}} \ J^\mu
\label{gope}\ee
where $\sqrt{g_{\rm bdy}}$ is the square-root of the induced metric on the boundary  and the current $J^\mu$ is defined by
\be
J^\mu =  \lim_{y\rightarrow x}\ \frac{1}{2} {\rm tr} \left< y \right| \hat{\gamma}^5 \hat{\gamma}^\mu \frac {\hat{\gamma}\cdot\nabla} {\left(- (\hat{\gamma}\cdot\nabla)^2 + m^2\right)}\left| x \right>
\ee
The two contributions in \eqn{gope} are typical for index theorems on manifolds with boundary. (See, for example, \cite{gpope}, for a discussion of index theorems for gravitational instantons. A similar structure is also seen in index theorems for Yang-Mills-Dirac operators on ${\bf R}^3\times {\bf S}^1$ \cite{poppitz}.)

\para
So far our discussion has been for a general anti-self-dual metric. At this point we restrict to the multi-Taub NUT spaces of interest, with metric given in \eqn{tn}. They have Pontryagin class
\be  \frac{1}{ 16\pi^2} \int d^4 x \sqrt{g} \,^\star {\cal R}_{\mu\nu\rho\sigma}{\cal R}^{\mu\nu\rho\sigma} =- 2k\label{pontytn}\ee
To compute the boundary in \eqn{gope}, we use some standard machinery \cite{erick}. 
The current is a local response to a nearby (as $x\rightarrow y$) excitation and its flux through the boundary can be computed using only the asymptotic form of the metric \eqn{tn}. Since the volume scales as $r^2$, anything decaying as $1/r^3$ or faster in the current vanishes. Accordingly, if we expand the covariant derivatives as $\nabla_\mu = \partial_\mu + \frac 1 2 t^{ab} \omega_{ab\mu}$, we have
\be
J^\mu = \frac 1 2 {\rm tr} \left< x \right| \hat{\gamma}^5 \hat{\gamma}^\mu \hat{\gamma}^\nu &&\!\!\left(\partial_\nu + \frac 1 2 \omega _{ab\nu} t^{ab}\right) \Bigg[ \frac 1 {\left(- \Delta_0 + m^2 \right) } \nn\\
&& + \frac {1} {\left(- \Delta_0 + m^2  \right)}  \omega_{ab}\,^\rho t^{ab} \partial_\rho \frac 1 {\left(- \Delta_0 + m^2  \right) } +\ldots \Bigg ] \left| x \right>
\nn \ee
The leading terms vanish using ${\rm tr} \hat{\gamma}^5 \hat{\gamma}^a \hat{\gamma}^b = 0$. Keeping only terms which survive asymptotically, we find
\be
J^\mu \longrightarrow \frac 1 2 {\rm tr} \left[ \hat{\gamma}^5 \hat{\gamma}^\mu \hat{\gamma}^\nu t^{ab} \right] \omega_{ab}\,^\rho \left< x \right|   \Bigg[ \frac{1}{2} \frac {g_{\nu\rho}} {\left(- \partial^2 + m^2 \right)  } +  \frac {\partial_\nu\partial_\rho} {\left(- \partial^2 + m^2  \right)^{2}}    \Bigg ] \left| x \right>
\nn \ee
The overall coefficient is determined by the trace of gamma matrices. It differs for spin-1/2 and spin-3/2:
\be
\frac 1 2 {\rm tr} \left[ \hat{\gamma}^5 \hat{\gamma}^\mu \hat{\gamma}^\nu t^{ab} \right] = \beta_s \epsilon^{\mu\nu a b} \ \ \ \ {\rm with}\ \ \ \  \beta_{1/2} = 1\ \ \ {\rm and}\ \ \  \beta_{3/2} = 4  \nn
\ee
Using the self-duality of the spin connection \eqn{sdspin}, we can then write
\be
J^\mu &\longrightarrow& -\beta_s\,\omega^{\mu\nu \rho} \left< x \right|   \Bigg[ \frac {g_{\nu\rho}} {\left(- \partial^2 + m^2 \right)  } +  \frac {2\partial_\nu\partial_\rho} {\left(- \partial^2 + m^2  \right)^{2}}    \Bigg ] \left| x \right>
\nn\\ &=&  -\beta_s\,\omega^{\mu\nu \rho} g^{-1/2} \frac{1}{2\pi L}\sum_n \int \frac{d^3 k}{(2\pi)^3}  \Bigg[ \frac {g_{\nu\rho}} {\left( k^2 + m^2 \right)  } -  \frac {2 k_\nu k_\rho} {\left( k^2 + m^2  \right)^2}   \Bigg ]
\nn \ee
where we have introduced a Fourier basis to integrate over the 4d momenta $k^\mu = ({\bf k},n/L)$. Our interest is in the outward flux, $J^i$ where we will take  $i=1,2,3$ to be  a tangent space index for simplicity. Asymptotically, the metric is locally flat and we  have $k^2 = {\bf k}^2 + n^2/L^2$. Using the explicit form of the spin connection, one finds that only the $\nu,\rho=4$ components contribute, and the relevant current is given by
\be
J^i &\longrightarrow& -  \frac{\beta_s}{2}  (\partial_i \log U)\,  \frac{1}{2\pi L}\sum_n\int \frac{d^3 k}{(2\pi)^3}  \Bigg[ \frac {1} {\left( {\bf k}^2 + n^2/L^2 + m^2 \right)  } -  \frac {2} {\left( {\bf k}^2 + n^2/L^2 + m^2  \right)^2} \frac{n^2}{L^2}  \Bigg ] \nn \\
&=&  + \frac{\beta_s}{2}  \partial_i\left( 1 + \frac{Lk}{2|{\bf x}|} \right) \, \frac{1}{8\pi^2 L} \sum_n \left[ \left(\frac{n^2}{L^2} + m^2 \right)^{1/2} + \left(\frac{n^2}{L^2} + m^2 \right)^{-1/2} \frac{n^2}{L^2} \right] \nn \\
&=& -\frac {\beta_s k} {32\pi^2 L} \, \frac {x^i}{|{\bf x}|^3}  \sum_n \left[ \left(n^2 + m^2L^2 \right)^{1/2} + \left(n^2 + m^2L^2 \right)^{-1/2} n^2 \right]
\nn \ee
where we have taken the liberty of regularising the linearly divergent term that appears in going from the first to the second line.
Finally, we need the fact that the asymptotic  flux is given by
\be
 \int dS_i \sqrt{g_{\rm bdy}}\,\frac{x^i}{|{\bf x}|^3} = 8 \pi^2 L
\nn\ee
Putting this together with the Taub-NUT Pontryagin class \eqn{pontytn}, the regularised index \eqn{gope} can be written as
\be
{\cal I}(m^2) =  -\frac{\alpha_sk}{12}  -\frac{\beta_s k}{4} \sum_{n\in {\bf Z}} \left[ \left(n^2 + m^2L^2 \right)^{1/2} + \left(n^2 + m^2L^2 \right)^{-1/2} n^2  \right]
\label{almosti}\ee

\subsubsection*{The Index}

Let us pause to compute the index of the Dirac operator in the multi-Taub-NUT backgrounds. As we saw previously, the index is given by ${\cal I}(m^2=0)$. In this limit, the sum above reduces to $4\zeta(-1) = -1/3$. (The sum is over both positive and negative integers which gives a factor of 2.) Combined with the contribution from the Pontryagin class, we find
\be {\cal I} = -\frac{k}{12} + \frac{k}{12} = 0\ \ \ \ \mbox{for  spin-1/2}\nn\ee
and 
\be  {\cal I} = +\frac{20k}{12} + \frac{4k}{12} = 2k\ \ \ \ \mbox{for  spin-3/2}\nn\ee
This agrees with the results of \cite{gpope}. This also confirms a statement that we made earlier: if we are interested in contributions to the superpotential, only the single Taub-NUT, with $k=1$, will play a role. Nonetheless, for completeness we will compute the determinants around an arbitrary multi-Taub-NUT background.

\para
It's instructive to return to the decomposition of spin-3/2 fermions in a self-dual background \eqn{gd1}. We see that the degrees of freedom include two anti-self-dual two-forms, $F^{\alpha\beta}$, transforming in the $(1,0)$ representation of $SO(4)$. These are the objects that  carry the zero modes.  The same objects appear in the decomposition of the metric \eqn{gd2} which  contains three anti-self-dual two forms   $H^{\alpha\beta}$.  This is the reason why the metric \eqn{tn} has $3k$ bosonic zero modes. These are identified with the positions ${\bf X}_a$ of the NUTs.

\subsubsection*{Back to the Determinants}

We now return to the task of computing the determinants. The sums in our expression \eqn{almosti} for ${\cal I}(m^2)$ are divergent. Although we have used zeta-function regularisation in the derivation of the first term in  \eqn{almosti}, at this stage it is important that we return to Pauli-Villars regularisation so that we can correctly match the finite terms with our one-loop counterterm \eqn{overthecounter}. We have 
\be  \log D(m^2) - \log D_0 &=& \int_0^1 \frac{d \lambda}{\lambda} \pv{I(\lambda m^2)} \nn\\ &=& -\frac{\beta_s k}{4} \sum_n \pv{ 2\sqrt{n^2 + m^2 L^2 } -2|n|-4|n|\log \left(\frac{1}{2} +\frac{1}{2} \sqrt{1 + \frac{m^2 L^2}{n^2}} \right) } \nn
\ee
where $\log D_0 = \lim_{\lambda\rightarrow 0}\, [\log  D(\lambda m^2)]_{\rm PV}$ is the logarithmic ratio of determinants in the limit in which all four fields in the Pauli-Villars regulator become massless.  The equality on the second line follows after noting that any $m^2$-independent piece in ${\cal I}(m^2)$ vanishes in the Pauli-Villars regulator.

\para
The sum above is now finite for each Pauli-Villars field individually. In the limit  $m^2\rightarrow 0$, the sum vanishes which means that it receives no contributions from the original field. But it still receives contributions from the three additional terms in the regularisation \eqn{pv}. Each of these has a large mass given by $\reg$ (or $\gamma\reg$ or $(\gamma-1)\reg + m^2$) and we are interested in the asymptotic form of the sum in the limit $\reg\rightarrow \infty$. We find that
\be
\int_0^1 \frac{d \lambda}{\lambda} \pv{I(\lambda m^2)} \longrightarrow \frac{\beta_sk}{12} \left[ \log(\mu^2 R^2) + C' \right] \nn
\ee
where $\mu^2 = (\gamma- 1)\reg^2/\gamma$ is the appropriate Pauli-Villars scale. The same quantity appeared in the one-loop counterterm \eqn{overthecounter}. The constant is given by $C'=-\log 4+1 - 24\zeta'(-1)$.

\para
The quantity $\log D_0 = \lim_{\lambda\rightarrow 0}\, [\log D(\lambda m^2)]_{\rm PV}$ is dominated by the zero modes. As we saw above, there are ${\cal I}=0$ zero-modes for spin-1/2 operators and ${\cal I} = 2k$ zero modes for spin-3/2. We have  
\be
D_0 = \frac{ (\lambda m^2)^{\cal I} (\lambda \gamma\reg^2)^{\cal I} } { (\lambda \reg ^2)^{\cal I} (\lambda (\gamma-1)\reg^2 + \lambda m^2)^{\cal I} } \longrightarrow \left(\frac{m^2}{\mu^2}\right)^{{\cal I}} \nn
\ee
We now have everything that we need to compute the one-loop determinants \eqn{newratio} about the $k$-centered Taub-NUT background.
%
%
%\be \Gamma  = \left(\frac{\det \Delta_{3/2\; +}}{\det \Delta_{3/2\; -}} \right)^{-1/2} \left(\frac{\det \Delta_{1/2\; +}}{\det \Delta_{1/2\; -}} \right)^{+1/4}\nn \ee
%
In the limit $m^2\rightarrow 0$, the determinants take the form
\be \Gamma = \left(\frac{\det \Delta_{3/2\; +}}{\det \Delta_{3/2\; -}} \right)^{-1/2} \left(\frac{\det \Delta_{1/2\; +}}{\det \Delta_{1/2\; -}} \right)^{+1/4} = m^{-2k}\,{\Gamma'}\nn\ee
which reflects the fact that $2k$ zero modes are carried by  $\Delta_{3/2\; +}$. The truncated determinants $\Gamma'$ are given by
\be 
\Gamma' = (\mu^2)^{41 k/48} \left(\frac{R^2}{A}\right)^{-7k/48}\label{atlast}\ee
where the constant numerical factor is 
\be A = 4e^{24\zeta'(-1)-1}\nn\ee
We note that we've seen the numbers that appear in \eqn{atlast} before . The fraction $41/48$ appeared as the beta-function for the running Gauss-Bonnet coupling \eqn{overthecounter}. This is not a coincidence. The fraction $7/48$ appeared in the one-loop shifted complex structure \eqn{ella}. This is not a coincidence either.

%\subsection*{The weird integral-sum result}
%
%This is obtained by (implicitly going back to) $\zeta$ regularising the summand, and then expanding it both for the large masses and the small mass separately. The small mass expansion one immediately finds vanishes for small masses $m$. This leaves only the large mass expansion. One finds that the only terms which do not decay in this limit give a contribution
%
%\be
%\int_0^1 \frac{d \chi}{\chi} \pv{I(\chi m^2)} &\sim& -\frac{\beta_s}{4} \sum_n \left[ 2 m + 2n(-1+ \log 4 - \log m^2R^2 )  \right]_{\rm {PV\ without\ light\ field}} \nn \\
%&=& \frac{\beta_s}{12} \left[ 1 + \log 4 - 24 \log A- \log m^2R^2 \right]_{\rm {PV\ without\ light\ field}} \nn \\
%&\sim& \frac{\beta_s}{12} \left( -1 - \log 4 + 24 \log A + \log \reg^2 R^2 - \log \gamma\reg^2 R^2  + \log (\gamma-1)\reg^2 R^2  \right) \nn \\
%&\sim& \frac{\beta_s}{12} \left( -1 - \log 4 + 24 \log A - \log \gamma  + \log (\gamma-1) + \log \reg^2 R^2 \right) \nn \\
%&\sim& \frac{\beta_s}{12} \left( C + \log \reg^2 R^2 \right) \nn
%\ee
%
%where $A$ is Glaisher's constant, and $C = -1 - \log 4 + 24 \log A - \log \gamma  + \log (\gamma-1)$.

\subsection{Zero Modes and Jacobians}\label{jacsec}

In any instanton computation, one should isolate the zero modes and replace their contribution to the path integral with a normal integration over the associated collective coordinates.
In doing so, we pick up a Jacobian factor for our troubles. For gravitational instantons, this procedure was described in \cite{gibperry}.

\subsubsection*{Bosonic Zero Modes}

We restrict our attention to the Taub-NUT metric \eqn{tn} with $k=1$. This metric has three collective coordinates which are identified with the position ${\bf X}$ of the nut. The three corresponding zero modes arise from translations and suitably gauge-fixed versions of them can be conveniently constructed by taking the Lie derivative of the metric along one of the three vector fields $\partial/\partial x^i$, $i=1,2,3$,
\be h_{\mu\nu}^{(i)} = {\cal L}_i g_{\mu\nu} = 2\nabla_{\mu}\nabla_{\nu}x^{(i)}\nn\ee
These zero modes are pure gauge. However, they arise from large gauge transformations which do not die off sufficiently fast at infinity and so should be thought of as physical. To see that they satisfy the transverse trace-free gauge condition, we use the facts that in our background we have
\be \nabla^2 x^{(i)} = g^{\mu\nu} \Gamma^{i}_{\mu\nu} = 0 \nn \ee
and also that we can commute certain derivatives through each other since ${\cal R}_{\mu\nu} = 0$. Consequently, we find
\be \nabla^\mu  \left(\nabla_\mu \nabla_\nu x^{(i)}\right) = \nabla_\nu \nabla^2 x^{(i)} = 0 \qquad\qquad g^{\mu\nu} \left(\nabla_\mu \nabla_\nu x^{(i)}\right) = \nabla^2 x^{(i)} = 0 \nn \ee
To compute the Jacobian, we need an inner product between the modes. This is inherited from the action and is given by,
\be
 \frac{M_{\rm pl}^2}{2}\int d^4x\sqrt{g}\ \frac{1}{2}h_{\mu\nu}^{(i)}h^{(j) \,\mu\nu} &=& M_{\rm pl}^2\int d^4x\sqrt{g} \left( \nabla_\mu \nabla_\nu  x^{(i)} \right)  \left( \nabla^\mu \nabla^\nu  x^{(j)} \right) \nn \\
&=& M_{\rm pl}^2\int d S^\mu \sqrt{g_{\rm bdy}} \left( \nabla_\mu \nabla_\nu  x^{(i)} \right)  \left( \nabla^\nu  x^{(j)} \right) \nn \\
&=& 2\pi M_{\rm pl}^2 L \int d^2 x \; e^{\mu}_k\, \frac{x^k}{r} r^2 \left( -\Gamma^i_{\mu\nu} \right)  \left( g^{\nu j } \right) \nn \\
&=& \pi M_{\rm pl}^2 L \int d^2 x \; \frac{x^k}{r} r^2 \left( - \delta^{ij} \partial_k - \delta^i_k \partial^j + \delta^j_k \partial^i \right) U \nn \\
&=& 2\pi^2 M_{\rm pl}^2 L^2 \; \delta^{ij} \nn
\ee
This we recognise as the Taub-NUT action, $S_{\rm TN} = 2\pi^2 M_{\rm pl}^2 L^2$. The upshot is that the integral over the three bosonic collective coordinates comes with the measure
\be \int d\mu_B = \int \frac{d^3X}{(2\pi)^{3/2}}\,S_{\rm TN}^{3/2} \label{measurebosonic} \ee

\subsubsection*{Fermionic Zero Modes}

As we saw in the previous section, the gravitino has two zero modes in the $k=1$ Taub-NUT background. These are Goldstino modes, arising from broken supersymmetry but, like their bosonic counterparts, are physical as they arise from large gauge transformations,  $\psi_\mu = \nabla_\mu \epsilon$. The $\epsilon$ parameter satisfies the gauge fixing condition
\be \gamma^\mu \psi_\mu = \Dslash \epsilon = 0  \nn \ee
The gravitino introduced in the original action \eqn{readyset} is a Majorana fermion. However, there is no Majorana condition in Euclidean space and, for this reason, it's simplest to work with a two component Weyl spinor formalism where
\be \psi_\mu = \left(\begin{array}{c} \psi_{\mu\alpha} \\ \bar{\psi}_\mu^{\ \dot{\alpha}}\end{array}\right)\nn\ee
The zero mode for this two-component spinor is then  $\psi_{\mu\alpha} = \nabla_\mu \epsilon_\alpha$, $\alpha=1,2$, and the zero mode equation reduces to
%
%\be \bar{\sigma}^{\mu} \nabla_\mu \chi = -\frac{i}{U} \; \sigma^i \partial_i \left( U^{1/2} \chi \right)  = 0  \nn \ee
%
%
\be \bar{\sigma}^{\mu} \nabla_\mu \epsilon = -i  \sigma^i\, \frac{\partial_i \left( U^{1/2} \epsilon \right)}{U}  = 0  \nn \ee
which has normalisable solutions of the form
\be \epsilon = \frac{1}{U^{1/2}} \xi \nn \ee
for any  constant spinor $\xi_\alpha$. (These are not to be confused with the right-handed spinors $\xi_{\dot{\alpha}}$ introduced in Section \ref{asdbacksec} which are associated to the unbroken supersymmetry. In contrast, the left-handed spinors $\xi_\alpha$ are associated to the broken supersymmetry.)

\para
The fermionic zero modes are accompanied by the measure
\be \int d\mu_F = \int d^2\xi\ {\cal J}_F^{-1}\nn\ee
The fermionic Jacobian, ${\cal J}_F$, is given by the overlap of zero modes,
\be
{\cal J}_F&=& \frac{M_{\rm pl}^2}{2} \int d^4 x d^2\xi\ \sqrt{g} \left( \nabla^\mu \epsilon \right)^\alpha\left( \nabla_\mu \epsilon \right)_\alpha \nn\\ &=& \frac{M_{\rm pl}^2}{2} \int dS^\mu d^2\xi\  \sqrt{g_{\rm bdy}} \,\epsilon^\alpha \left( \nabla_\mu \epsilon \right)_\alpha \nn \\
&=&  \pi M_{\rm pl}^2 L \int d^2x \ x^i\, r \, (\partial_i U^{-1/2}) =\frac{1}{2}S_{\rm TN}\nn\ee
where, in the last line, we use the normalisation $\int d^2\xi\ \xi^2=1$.
%&=&  \pi^2 M_{\rm pl}^2 L^2 \left( f^\star f \right) \nn \\
%&=&  \frac 1 2 S_{\rm TN} \left( f^\star f \right)
%
%There are various issues here. For instance, it is a true fact that $\chi \sigma^m \bar{\sigma}^n \psi = \psi \sigma^n \bar{\sigma}^m \chi$ for anticommuting spinors and therefore $\lambda \sigma^{[m} \bar{\sigma}^{n]} \lambda = 0$. {\bf However I've dropped something slightly different in moving to the non-covariant derivative. I also don't really know what I'm doing with that star actually.}
%
%\para
%Now the zero mode integral is over two Grassman parameters for the two independent choices of constant spinor $f$: let
%
%\be \psi_\mu = \nabla_\mu\left( \frac{f}{U^{1/2}} \right) \nn \ee
%
%Then the fermionic sector contributes
%
%\be \int d^2 f \frac{2}{S_{\rm TN}} \label{measureferm} \ee
%
\para
Putting this together with the bosonic measure \eqn{measurebosonic}, we find that the integration over all collective coordinates is accompanied by the Jacobian factor
\be
\int d\mu_B d\mu_F = \int \frac{d^3X}{(2\pi)^{3/2}} \int d^2 \xi \ 2S_{\rm TN}^{1/2} \label{measurezero} \ee

\subsection{Computing the Superpotential}\label{finalsec}

We now have all the ingredients necessary to compute the instanton-generated superpotential. We start by computing the two-point function of the 3d spin-1/2 fermion $\chi$ which arises under dimensional reduction \eqn{fermidimred} from $\psi_4$. As we have just seen, in the background of Taub-NUT we can turn on a fermionic zero mode. For $\chi$, this is given by
\be \chi_\alpha = \frac{1}{2}\omega_{ab4} \left(\sigma^{ab}\xi\right)_\alpha = \frac{\partial_i U}{U^{3/2}}\,\left(\sigma^{i4}\xi\right)_\alpha\nn\ee
Far from the NUT itself, the zero mode becomes
\be \chi_\alpha \rightarrow \pi L S_F(x-X)_{\alpha}^{\ \beta}\xi_\beta\nn\ee
where $S_F(x) = \gamma_{\rm 3d}^ix_i/4\pi x^3$ is the flat-space propagator. This form will suffice for our instanton computation.
Using our results for the action \eqn{wtn}, the one-loop determinants \eqn{atlast} and  the measure \eqn{measurezero}, we have the two-point function
\be \langle \chi_\alpha(x)\chi_\beta(y)\rangle = \int \frac{d^3X}{(2\pi)^{3/2}} \int d^2 \xi  \!&&2S_{\rm TN}^{1/2}\ \mu^{41 /24} \left(\frac{R^2}{A}\right)^{-7/48} \!\!\!e^{-2\pi^2 M_{\rm pl}^2 R^2 + i\sigma} e^{-\tau^\star_{\rm grav}}  \nn\\  &&\ \ \ \ \ \ \  \times\ \pi^2 L^2 S_F(x-X)_\alpha^{\ \gamma}\xi_\gamma\, S_F(y-X)_\beta^{\ \delta}\xi_\delta\nn\ee
%&&  \times\ \frac{\partial_i U \partial_j U}{U^3}\,(\sigma^{i4}\xi)_\alpha (\sigma^{j4}\xi)_\beta\nn\ee
%
Let's firstly explain why the various fractions that appear in the determinants are not coincidental. The power of the Pauli-Villars scale $\mu^{41/24}$ combines with the $e^{-\tau^\star_{\rm grav}}$ factor to give rise to the RG-invariant scale that we introduced in \eqn{lambda},
\be (\Lambda^\star_{\rm grav})^{41/24}  = \mu^{41/24}  e^{-\alpha(\mu)+2i\theta}\nn\ee
As we explained in Section \ref{classec}, the complexified $\Lambda_{\rm grav}$ sits in a chiral multiplet and so can appear in the superpotential. Meanwhile, the power of $(R^2)^{-7/48}$ combines with the instanton action  to give $e^{-{\cal S}^\star}$ where ${\cal S}$ is the one-loop corrected complex structure introduced in \eqn{ella},
\be {\cal S} = 2\pi^2 M_{\rm pl}^2 R^2 + \frac{7}{48}\log(M^2_{\rm pl}R^2) +i\sigma \nn\ee
Once again, ${\cal S}$ is the lowest component of a chiral multiplet and so can naturally appear in a superpotential. (There are further powers of $R$ buried in the factor $S_{\rm TN}^{1/2}$ in the two-point function but, as we will now see, these do not appear in the superpotential.)

\para
Continuing with the computation, we have
\be  \langle \chi_\alpha(x)\chi_\beta(y)\rangle = \frac{A^{7/48}}{2(2\pi)^{3/2}} \left(\frac{\Lambda^\star_{\rm grav}}{M_{\rm pl}}\right)^{41/24}\! \!S_{\rm TN}^{3/2}\, e^{-{\cal S}^\star} \int {d^3X}\ 
S_F(x-X)_{\alpha\gamma}S_F(y-X)_{\beta\delta}\epsilon^{\gamma\delta}\nn\ee
%
%
%\be  \langle \chi_\alpha(x)\chi_\beta(y)\rangle = A^{7/48} \left(\frac{\Lambda_{\rm grav}}{M_{\rm pl}}\right)^{41/24}\! \!S_{\rm TN}^{3/2}\, e^{-{\cal S}} \int \frac{d^3X}{(2\pi)^{3/2}} \int d^2 \xi \ 2S_{\rm TN}^{1/2}\ \frac{\partial_i U \partial_j U}{U^3}\,(\sigma^{i4}\xi)_\alpha (\sigma^{j4}\xi)_\beta\nn\ee
%
We want to write down a low-energy effective action for $\chi$ which captures this two-point vertex. This can be simply done if the kinetic term \eqn{chikin} around a flat background is supplemented by the interaction term 
\be S_{\chi} =   \int d^3x\,\sqrt{-g_{(3)}}\ M_3 \left[\bar{\chi}\!\delslash\chi  + \frac{M_3A^{7/48}}{4(2\pi)^{3/2}} \left(\frac{\Lambda_{\rm grav}}{M_{\rm pl}}\right)^{41/24}\! \!S_{\rm TN}^{3/2}\, e^{-{\cal S}}\, \chi\chi + {\rm h.c.} \right]\ \ \ \ \ \ \label{gettingthere}\ee
where we're now working in the choice of coordinates of \eqn{tn} such that $R(x)\rightarrow L$ asymptotically.
We would like to determine the supersymmetric completion of this interaction term.

\subsubsection*{Supersymmetric Effective Action}

The spin-1/2 fermion $\chi$ is related to the superpartner of our complex scalar ${\cal S}$ defined classically by \eqn{complexclas}. However, there is an important normalisation that must  be determined. We denote by $\Psi$ the spin-1/2 Dirac fermion that sits in the chiral multiplet with ${\cal S}$. By supersymmetry, the kinetic term for $\Psi$ must agree with that of ${\cal S}$ in \eqn{sact}, namely
\be S_{\Psi} =  M_3\int d^3x\,\sqrt{-g_{(3)}}\ \frac{1}{({\cal S}+{\cal S}^\dagger)^2}\bar{\Psi}\!\Dslash\Psi\nn\ee
Restricting to a flat background, and comparing to \eqn{chikin}, we learn that the correctly normalised superpartner of ${\cal S}$ is given by
\be \Psi = 2\pi M_3 R \chi\nn\ee
The instanton-generated $\Psi\Psi$ vertex in the low-energy effective action arises from a superpotential. The general form involves a number of terms. (See, for example, \cite{wb} for the general form in four-dimensions, or \cite{dewitt} for the three-dimensional effective action.) However, to the order that we're working, only the leading term contributes and the fermionic part of the action should take the form
\be
S_{\Psi} =\int d^3x \sqrt{g_{(3)}} \,  M_3\left[ (\partial \bar{\partial} K)\,  \bar{\Psi} \Dslash \Psi  +   \frac{1}{2}(e^{K/2}\partial \partial {\cal W}) \, \Psi \Psi + {\rm h.c.} \right] \label{gotthere}
\ee
where for the purposes of this calculation it suffices to use the classical K\"ahler potential $K=-\log({\cal S}+{\cal S}^\dagger)$ defined in \eqn{kahler}. Comparing the two expressions \eqn{gettingthere} and \eqn{gotthere}, we find that the superpotential is given by,
\be {\cal W}   = C M_3 \left(\frac{\Lambda_{\rm grav}}{M_{\rm pl}}\right)^{41/24} e^{-{\cal S}}\nn\ee
with the overall constant 
\be C = \frac{\left(4e^{24\zeta'(-1)-1}\right)^{7/48} }{ 2(4\pi)^{3/2}}\nn\ee
Note that the superpotential is not invariant under the $U(1)_J$ symmetry which shifts the dual photon. Further, the Yukawa vertex in \eqn{gotthere} explicitly breaks the $U(1)_R$ symmetry under which the gravitino is charge; this is manifestation of the axial anomaly \eqn{anomaly}. However, a combination of $U(1)_J$ and $U(1)_R$ symmetry survive.

\subsubsection*{The Potential}

The supersymmetric completion of the Yukawa term is a potential. In three-dimensional supergravity, this is given by (see, for example, \cite{dewitt,grimm})
\be V = M_3\,e^{K}\left( (\partial\bar{\partial}{ K})^{-1}\,|D{\cal W}|^2 - 4|{\cal W}|^2\right)\nn\ee
with $D{\cal W} = \partial {\cal W} + (\partial K){\cal W}$. This potential includes some critical points at ${\cal S}\sim {\cal O}(1)$. They are not to be trusted as they lie outside the semi-classical regime  of large ${\cal S}$ where we performed our calculation. Instead,  at large ${\cal S}$, the potential is dominated by the $|{\cal W}'|^2$ term and takes the runaway form
%
%\be V \sim M_3 \,({\cal S}+{\cal S}^\dagger) \,e^{-({\cal S}+{\cal S}^\dagger)}\nn\ee
%
\be V \sim M_3^3 (R M_{\rm pl})^{-2} (R \Lambda_{\rm grav})^{41/12}\exp\left(-4\pi^2M_{\rm pl}^2R^2\right)\nn\ee
We learn that the Kaluza-Klein compactification of ${\cal N}=1$ supergravity on ${\bf R}^3\times {\bf S}^1$ is not a ground state of the theory. This instanton-generated potential causes the circle to decompactify to large radius $R$.

\newpage
\section*{Acknowledgements}

This project was born out of discussions with Sean Hartnoll and we're grateful for his input and collaboration at this early stage. We would also like to thank  Dan Butter, Nick Dorey, Michael Green, Gary Gibbons, Stephen Hawking, Shamit Kachru, Sameer Murthy, Malcolm Perry, Chris Pope, Nati Seiberg, Harvey Reall, Stefan Vandoren and Kenny Wong for many useful and detailed conversations, and Nava Gaddam for pointing out a typo in an earlier version of the paper. We are supported by STFC and by the European Research Council under the European Union's Seventh Framework Programme (FP7/2007-2013), ERC Grant agreement STG 279943, Strongly Coupled Systems.

\end{document}